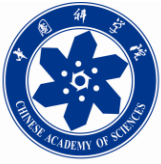

**中国科学院大学**
**University of Chinese Academy of Sciences**

# 硕士学位论文

## Estimation of land surface Evapotranspiration in Nepal using Landsat based METRIC model

### 利用 METRIC 模型和 Landsat 资料对尼泊尔地区蒸散发估算研究

作者姓名：**Shailaja Wasti**

指导教师：马伟强 研究员

学位类别：理学硕士

学科专业：大气物理学与大气环境

研 究 所：中国科学院青藏高原研究所

答辩委员会主席：＿＿＿＿＿＿

**2020 年 5 月**

# Estimation of land surface Evapotranspiration in Nepal using Landsat based METRIC model

**By**

**Shailaja Wasti**

**A Dissertation Submitted to**

**University of Chinese Academy of Sciences**

**In partial fulfillment of the requirement**

**For the degree of**

**Master of Atmospheric Physics and Atmospheric Environment**

**Institute of Tibetan Plateau Research, Chinese Academy of Sciences**

**May, 2020**

## Abstract

Nepal has a topographically diverse terrain with variations from plain flatland to high hills and mountains within a small region. Due to the lack of abundant ground-based measurement sites, it is difficult to estimate evapotranspiration (ET) covering most of the areas of Nepal from lower to higher elevations. In this study, we proposed to use a remote sensing-based METRIC (Mapping Evapotranspiration at high Resolution with Internalized Calibration) model for estimating ET in Nepal. Landsat 8 imagery, which provides a fine spatial resolution (30 m), was used for the estimation. The results obtained from the METRIC model were compared with ground-based measurements from an independent Eddy-Covariance (EC) station. Besides ET, the estimated surface temperatures ($T_s$) from the remote sensing model were also compared with ground-based measurements for model validation. The results obtained from the remote sensing model were close to the ground-based measurements which establish the accuracy of the model. Root mean square error (RMSE) for hourly and daily ET was obtained as 0.06 mm/hr and 1.24 mm/day while mean bias error (MBE) for hourly and daily ET was observed as 0.03 mm/hr and 0.29 mm/day, respectively. RMSE and MBE for $T_s$ were obtained as 4.93 ˚C and -0.49 ˚C, respectively.

Further, we analyzed the variations of ET with elevation for six different months and found that ET was inversely related to elevation, in general, over the regions of Nepal. $T_s$ and vegetation were generally higher in the area where ET is high. Therefore, an inverse relation of ET and elevation was observed because $T_s$ and vegetation decrease with increasing elevation.

To the best of our knowledge, our study is a first that has investigated ET estimation over a topographically diverse region of Nepal with fine spatial resolution afforded by Landsat 8. Such large-scale ET estimation in Nepal has several applications. This can be leveraged for agricultural planning and forecasting in a country where most of the population still relies on agriculture for their daily livelihood.

## 摘要：

尼泊尔地区地势复杂多样，有平原，小丘陵和山区地形。由于缺乏大量的地面观测站点，因此很难从低海拔到高海拔估算和验证尼泊尔大部分地区的蒸散量（ET）。在本研究中，我们使用了基于遥感的 METRIC（计算高分辨率的 ET 数据且经过内部校正校准）模型来估计尼泊尔的 ET。鉴于 Landsat 8 数据有较好的空间分辨率（30m），可用于估算较为准确的 ET。从 METRIC 模型计算获得的结果与当地架设的涡动相关系统（EC）站的地面测量结果进行了比较。除了估算 ET 以外，我们还将遥感模型所估算的地表温度（$T_s$）与基于地面的观测值进行比较，以进行模型验证。结果表明遥感模型所获得的结果接近于地面实际观测值，从而确定了模型的准确性。其中逐小时和逐天 ET 的均方根误差（RMSE）分别为 0.06mm/hr 和 1.24mm/day，而逐小时和逐天 ET 的均方误差（MBE）分别为 0.03 mm/hr 和 0.29 mm/day。地表温度的 RMSE 和 MBE 分别为 4.93℃ 和-0.49℃。

此外，我们分析了尼泊尔不同地区六个月的 ET 随海拔的变化特征，结果表明 ET 与海拔呈负相关关系。由于 ET 较高的地区地表温度和植被覆盖率通常较高，而地表温度和植被覆盖会随着海拔的升高而降低，因此观测到 ET 与海拔呈现出反比的关系。

我们的研究首次利用了具有高空间分辨率的 Landsat 8 对尼泊尔不同地形区域的 ET 进行了估算及研究。这种在尼泊尔的大规模 ET 估算具有多种应用。 如，它可以进行适当的农业规划和预测，而这对于一个大多数人口仍然依靠农业维持其日常生计的国家而言非常重要。

**关键词**：蒸散发量，METRIC 模型，Landsat 8，海拔高度, 复杂的地形

**Table of Contents**

## List of figures

**List of tables**

# List of Acronyms

| | |
|---|---|
| DEM | Digital Elevation Model |
| DN | Digital Number |
| DOY | Day of the Year |
| EC | Eddy Covariance |
| ET | Evapotranspiration |
| $ET_{inst}$ | Instantaneous ET |
| G | Soil Heat Flux |
| H | Sensible Heat Flux |
| K | Kelvin |
| Km | kilometer |
| LAI | Leaf Area Index |
| LE | Latent Heat Flux |
| m | Meter |
| MBE | Mean Bias Error |
| METRIC | Mapping Evapotranspiration at high Resolution with Internalized Calibration |
| MODIS | Moderate Resolution Imaging Spectroradiometer |
| NASA | National Aeronautics and Space Administration |
| NDVI | Normalized Difference Vegetation Index |
| NIR | Near Infrared |
| OLI | Operational Land Imager |
| R | Red |
| $R_n$ | Net Radiation |
| RMSE | Root Mean Square Error |
| SAVI | Soil Adjusted Vegetation Index |
| SEB | Surface Energy balance |
| $T_s$ | Surface Temperature |
| TB | Temperature Brightness |

| TIRS | Thermal Infrared Sensor |
| TOA | Top of Atmosphere |
| USGS | United States Geological Survey |

# Chapter 1 Introduction

## 1.1 Background

Solar energy is the main source of energy for most atmospheric dynamics (Dungey, 1961). Interchanges of the solar energy within the climatic system and their exchanges with outer space balance the global energy system (Wild et al., 2013). Starting with the incoming solar radiation as the fundamental source of energy to run the earth system, there is a balance of incoming and outgoing energy at the top of the atmosphere. Some portion of the incoming solar radiation is absorbed by the earth's surface and is referred to as the soil heat flux. Another portion transfers back to the atmosphere from the soil as conductive heat transfer and is referred to as sensible heat flux. The remaining energy component is called the latent heat flux which is the energy component associated with evapotranspiration (ET), a process of evaporation and transpiration of a large volume of water to the atmosphere from the soil, water bodies, and vegetation.

ET, as the name suggests, consists of two components: evaporation and transpiration. Incoming solar radiation absorbed by land surfaces is used to evaporate water. Plants, soil, snow cover, and open bodies also absorb and re-emit a portion of radiant energy as latent heat, and the associated water vapor loses to the atmosphere through evaporation. Similarly, green plants lose water vapor to the atmosphere from stomatal pores through the process of transpiration (Zhang et al., 2016a). ET plays a major role in the climate dynamics in a region; ET has an impact on land surface temperature regulation through involvement in the surface energy balance. ET varies from region to region depending upon several factors. ET of an area is dependent on weather conditions like air temperature, humidity, wind speed, solar radiation, plant and soil characteristics, etc. (Allen et al., 1998; George et al., 2002; Valipour, 2017).

ET is an important part of both the water and the energy cycle. It is through this mechanism of ET that the relation between the energy and water cycle is established via land-atmosphere interaction (Jung et al., 2010). The schematic diagram for the energy flow of the global energy balance system is shown in **Figure 1**.

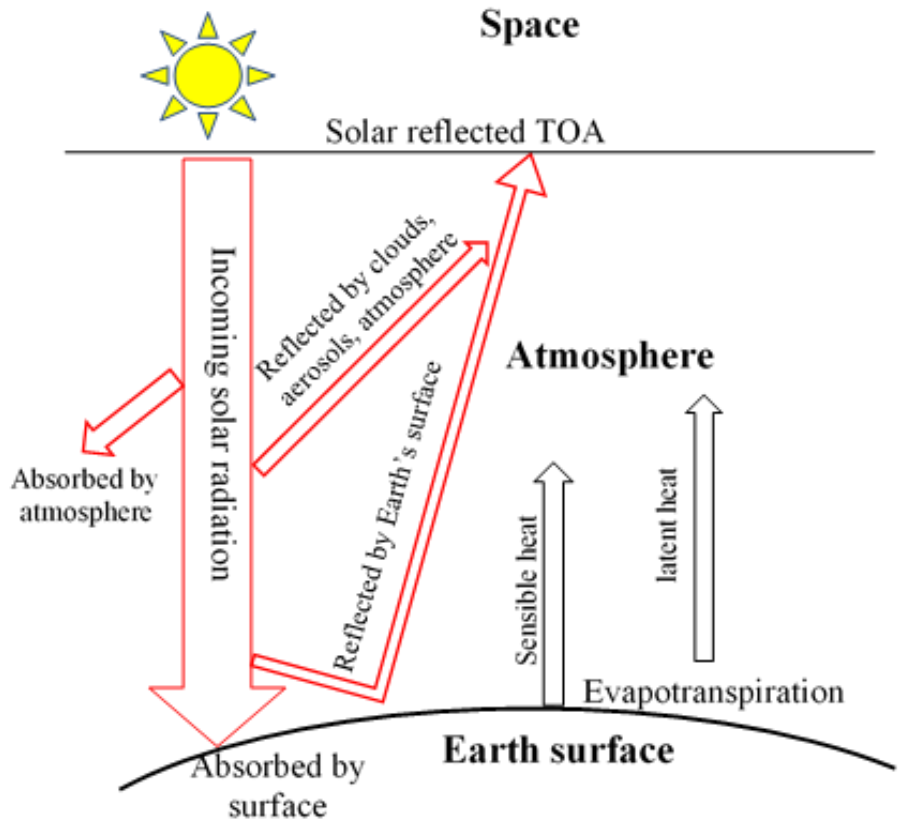

**Figure 1** Schematic diagram of the fundamental energy flow for the global energy balance system of the earth

ET computation and monitoring have several applications. A portion of water from the land surface loses due to ET that then influences soil moisture, agricultural drought, and hydrological drought. ET thus subsequently becomes one of the key variables for applications such as climate prediction, water resources management, and agricultural food production (Fisher et al., 2017; Goulden and Bales, 2014; Meng et al., 2014). It should be taken into important consideration when assessing response in the presence of large uncertainty in climate (Friedlingstein et al., 2014; Mancosu et al., 2015). Accurate estimation of ET can help to maximize crop productions with a minimum water loss and also helps in minimizing management costs (Wagle and Gowda, 2019).

Besides the application areas like agriculture, ET itself is an important part of the water cycle and climate regulation cycle. Large volumes of water from the soil and vegetation transfers to the atmosphere by ET. Globally, ET returns about 60% of annual land precipitation to the atmosphere (Koster et al., 2004). However, the contribution of ET to transfer water to the atmosphere has regional variations. But even then, in typical regional areas, the contribution from ET is more than half of the total precipitation and is almost equal to the precipitation in the semi-arid region (Engman and Gurney, 1991). Thus ET affects precipitation and the latent heat flux associated with ET helps to control surface temperature (Koster et al., 2004; Seneviratne et al., 2006; Vautard et al., 2007). By the use of a biophysical process-based model, the researchers found that within the global ET, transpiration ($E_T$) constitutes about 52%, soil evaporation ($E_S$) constitutes 28% and canopy evaporation ($E_C$) constitutes 20% (Choudhury and DiGirolamo, 1998). In another estimate of 2005 from Global Wetness Project 2 (Dirmeyer et al., 2006), the composition of global ET was 48% from $E_T$, 36% from $E_S$ and 16% from $E_C$. The global land surface ET distribution for 2006 which was obtained from multi-satellite information is shown in **Figure 2**, as an example of ET distribution over different regions of the world.

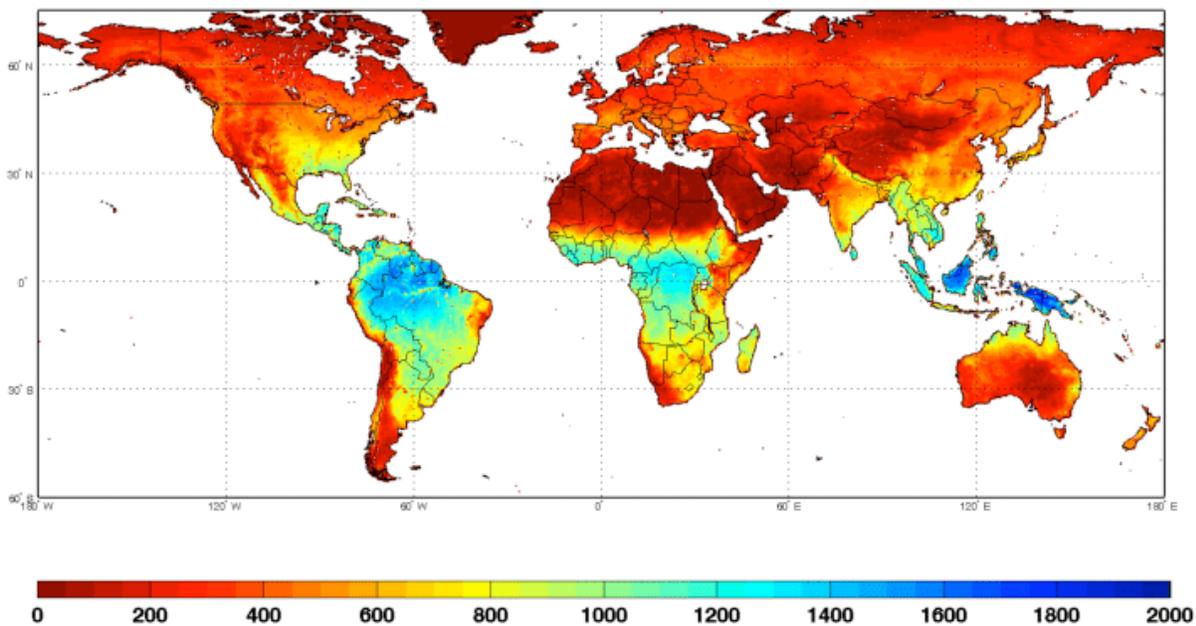

**Figure 2** Global land surface ET distribution for the year 2006 (Miralles et al., 2011)

Many empirical methods have been developed to estimate ET using climatic and meteorological variables. The eddy covariance technique, energy balance Bowen ratio, weighing lysimeter, pan-measurement, sap flow, scintillometer, etc. are some of the classical approaches for estimating ET based on field measurements (Idso et al., 1975). However, it is often difficult to measure ET with the ground station for large topographically diverse areas. Therefore, various remote sensing-based surface energy balance models have also been developed in the past years for ET estimation on a large scale.

Over the years, several models have been developed for ET estimation, each suited for different landscape types (Kalma et al., 2008). Surface energy balance index (SEBI) (Menenti, 1993), surface energy balance algorithm for land (SEBAL) (Bastiaanssen et al., 1998b), simplified-surface energy balance index (S-SEBI) (Roerink et al., 2000), surface energy balance system (SEBS) (Su, 2002), and METRIC (mapping evapotranspiration at high resolution and with internalized calibration) (Allen et al., 2007) models are some examples of surface energy balance model based on energy balance closure. In these models, ET is estimated as a residual concept through several surface properties such as albedo, leaf area index, vegetation cover, and surface temperature (Norman and Becker, 1995). Further introduction about these different models for ET estimation will be given in Chapter 1.3. Next, we introduce remote sensing system for the estimation of climate parameters such as ET.

## 1.2 Remote sensing system

Remote sensing is the observation technology where satellite-based sensors are utilized for measuring earth's surface features using reflected and emitted energy from the earth's surface. Representing one of the first advances in remote sensing, National Aeronautics and Space Administration (NASA) launched the first space-based meteorological satellite named Television Infrared Observation Satellite (TIORS) in 1960. This satellite was designed especially for weather forecasting. Later in 1972, NASA launched the first Landsat satellite which provides a high-resolution remote sensing system. Landsat 1 had two sensors onboard i.e. Return Beam Vidicon (RBV) and Multispectral Scanner System (MSS) that gives imagery in both digital and multispectral form. In 1991, European Space Agency's first Earth observing satellite program launched its first satellite named European Remote Sensing (ERS-1) and in 1995 they launched ERS-2. These satellites have special earth observation instrument called a Synthetic Aperture Rader (SAR) which produces a high-resolution image.

From 1972 onwards, Landsat satellites and many other satellites have been launched for remote sensing purposes. There are several satellite sensors developed for remoted sensing: Advanced Spaceborne Thermal Emission and Reflection Radiometer (ASTER), Advanced Very High-Resolution Radiometer (AVHRR), Landsat Thematic Mapper (TM), Landsat Enhanced Thematic Mapper+ (ETM+), Landsat Operational Land Imager (OLI) and Thermal Infrared Sensor (TIRS), Advanced Microwave Scanning Radiometer-Earth Observing System (AMSR-E), Moderate Resolution Imaging Spectro-radiometer (MODIS) and, many more which have spatial resolution of 30 m to few kilometers. A summary of these different satellite sensors is provided in **Table 1**. These sensors collect the data from the earth's surface at different wavelengths. The collected data are first corrected to radiometric and geometric correction and then used to analyze different features of the earth using climatic models. Estimation of ET based on remote sensing is also mostly based on models to map remotely sensed image data to ET distribution at the earth's surface. Next, we provide a detailed description of different energy balance-based models used for ET estimation from a remote sensing system.

**Table 1** Different sensors with their resolution and date of operation

| Sensor | Spatial Resolution | Temporal Resolution | Year of Operation |
|---|---|---|---|
| AVHRR | 1.1 km | 14 days | October 13, 1978 |
| Landsat TM | 30 m | 16 days | July 16, 1982 |
| Landsat ETM+ | 30 m | 16 days | April 15, 1999 |
| Landsat OLI and TIRS | 30 m | 16 days | February 13, 2013 |
| ASTER | 15 m to 90 m | 16 days | December 18, 1999 |
| MODIS | 250 m to 1 km | 1 to 2 days | December 18, 1999 |
| AMSR-E | 25 km | 5 days | May 02, 2002 |
| Sentinel-1A | 20 m | 12 days | April 03, 2014 |

## 1.3 Methodological overview of different energy balance methods

Remote sensing-based ET estimation has become a popular tool in the past few decades due to its wide and continuous coverage, and reasonably favorable accuracy. Many remote sensing-based surface energy balance (SEB) models have been developed to estimate ET in large spatial scales which differ with respect to landscape type and spatial extent of model application, type of remote sensing data, and required meteorological and land-cover data (Kalma et al., 2008). Each SEB model is based on the fundamental energy balance equation which is expressed as:

$$R_n = G + H + LE \qquad\qquad (1.1)$$

where $R_n$ is the net radiation, G is the soil heat flux, H is the sensible heat flux and LE is the latent heat flux consumed by ET. All the fluxes are expressed in watts per meter square (Wm-2).

Net radiation represents the total radiation at the earth's surface with incoming solar radiation as the source of energy. Net radiation is obtained by adding the incoming long and short-wave solar radiation and deducting the outgoing radiation components from the sum. These outgoing radiation components are represented as a fraction of incoming components and depend on surface properties like surface thermal emissivity, among others.

When the earth's surface receives radiation energy, some components of this radiation is stored in the soil and plants covers owing to heat conduction. This stored energy component is referred to as the soil heat flux. Soil heat flux then can also be represented as a fraction of net radiation at the earth's surface and depends on the surface properties like the vegetation present in each region. A component of heat flux at the earth's surface is also conducted to the atmosphere through conduction. This is represented by the sensible heat flux. Latent heat flux, which is the energy component associated with ET, is the heat component associated with evaporation from land surface and transpiration from the plant surface.

Different surface energy balance models differ in how different heat fluxes are computed. Here we discuss some of the remote sensing based SEB models that build background for our selected METRIC based model. The schematic view of the procedure for the overall SEB model is shown in **Figure 3**.

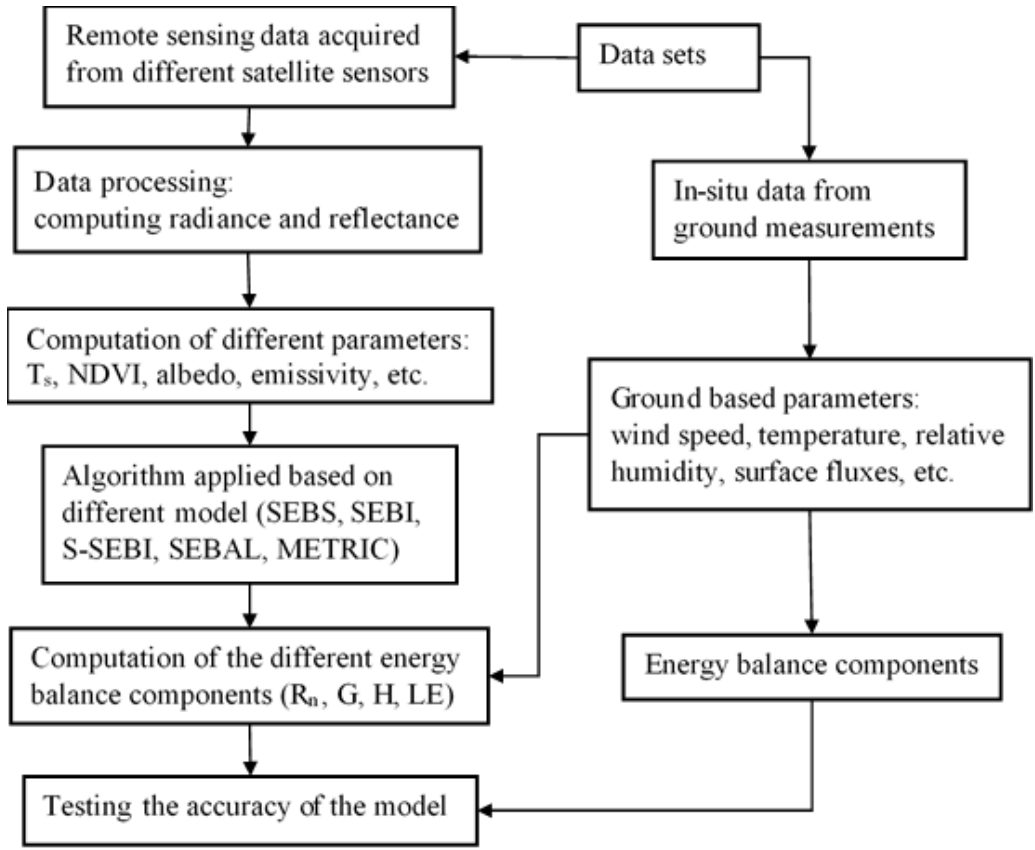

**Figure 3** Flow chart of the generalized SEB model for ET estimation using remote sensing and in-situ data

**Surface energy balance index (SEBI) model**

SEBI is a single-source energy balance model proposed by Menenti (1993) to estimate the ET from evaporative fraction based on the contrast between dry and wet regions. The dry region in this model is assumed to have zero surface ET while the wet region in this model assumes to have high ET. This model relates the effects of surface temperature and aerodynamics resistance directly to ET. This model is based on satellite measurement data, but also ground-based measurements are needed.

**Surface energy balance system (SEBS) model**

SEBS is a modified form of the SEBI model. Su (2002) developed this model which is based on the surface energy balance system model for the estimation of land surface energy balance using remotely sensed data. In this model also, dry limit (corresponding to dry region) of the LE is zero and at the wet limit (corresponding to wet region) ET is assumed to be maximum. The main advantages of SEBS over SEBI are uncertainty from $T_s$ and meteorological variables are limited in this model and also roughness height for heat transfer is computed instead of using fixed values. The demerits of this model are that it requires ground-based parameters which makes it difficult to compute ET in the places which lack in-situ measurement sites.

**Simplified surface energy balance index (S-SEBI) model**

For the estimation of the surface flux from remote sensing, a simplified method was derived from SEBI called a simplified surface energy balance index by Roerink et al. (2000). The main assumption of this model is that the evaporative fraction varies linearly with $T_s$ for a given surface albedo. This model assumes that maximum $T_s$ corresponds to the minimum LE and minimum $T_s$ corresponds to the maximum LE. This model has both merits as well as demerits. The main advantage of this model is that it doesn't require ground-based measurements and the disadvantage is that the extreme temperatures are location specific.

**Surface energy balance algorithm for land (SEBAL) model**

SEBAL developed by Bastiaanssen et al. (1998b) is an image-processing model for estimating ET, with minimum ground-based measurement required, as a residual of the surface energy balance. The major advantages of SEBAL for estimation of ET are automatic internal calibration, a requirement of minimum ground-based measurements, and avoidance of needing exact atmospheric corrections. Despite its advantages, there are some shortcomings such as subjective specifications of representative hot/dry and wet/cold pixels within the image are required. Under several climatic conditions, SEBAL has been tested for both field and catchment scales in more than 30 countries with the typical accuracy at field scale being 85% for daily and 95% for seasonal scales (Bastiaanssen et al., 2005; Bastiaanssen, 2000).

**Mapping Evapotranspiration at high resolution and with internalized calibration (METRIC) model**

METRIC is an energy balance model that utilize high-resolution data, developed by Allen et al. (2007) to avoid the limitation of the SEBAL. It is an image processing tool for mapping regional

ET from remotely sensed data in the visible, near-infrared and thermal infrared spectral regions along with ground-based wind speed and near-surface dew point temperature. METRIC model is like the SEBAL model but surface slope, aspect, and temperature lapsing rate can be considered in the METRIC model which makes the model applicable for mountainous areas also. METRIC model has been widely used in different areas by utilizing high-resolution satellite imagery along with weather station data. Chavez et al. (2007) and other researchers (Folhes et al. (2009), Healey et al. (2011), Zhang et al. (2015a), and Liebert et al. (2016)) applied METRIC model to estimate ET in Texas high plains, Northeast of Brazil, semi-arid sandhills of Nebraska, Maui, Hawaii, and Lower Virgin River. Results from these all showed a strong relationship between observed and METRIC estimated ET indicating that the Metric model is useful for estimating accurate ET at local and field scales.

Next, we review some of the earlier works that have used remote sensing-based ET estimation in different areas of the world. Then we review those works that have particularly focused on ET estimation for Nepal.

## 1.4 ET estimation for different regions in the world

Several remote sensing-based studies have been carried out around the world on land surface ET. In this section, we discuss some of these studies to highlight the methods used by different authors and general observations in their ET estimation results. Mielnick et al. (2005) conducted research on Chihuahuan desert grassland by the Bowen ratio approach where the results showed that the daily ET rates were maximum when precipitation and radiation were high. In the multi-model decadal analysis by the second Global Soil Wetness Project (GSWP-2), it was observed that latent heat flux is higher in the tropical rain forests and subtropical forest regions whereas it is lower in deserts, high mountain regions and the polar zones (Dirmeyer et al., 2006). Using remote sensing-based method Tsouni et al. (2008) estimated daily actual ET in a wide irrigated area over Thessaly Plain in Greece. The results showed great agricultural importance of ET during the summer season. The authors concluded that the wind speed has a significant influence on ET estimates as its high-value results in high underestimates of ET.

ET was computed in Australia by the application of the SEBS model in earlier studies using Landsat-5 (Ma et al., 2012) and ASTER (Ma et al., 2013). The authors found that remote sensing-based ET estimation matched closely with field measurements (ground-based measurements). By utilizing remote sensing data, Han et al. (2017) used a SEBS model to see the annual trends (2001-2012) of the heat fluxes on the Tibetan Plateau and the results thus obtained shows increasing trends for the net radiation and the latent heat flux while it shows decreasing trends in case of the sensible heat flux. Song et al. (2017) used remote sensing and the meteorological data and modified Penman-Monteith based algorithm to estimate ET on the TP for a decade (2000 – 2010). The author obtained highest ET as 680.9 mm/year in open water bodies and lowest as 254.0 mm/year in open shrubland.

Lian and Huang (2016) compared three contextual remote sensing-based models (METRIC— Mapping Evapotranspiration at High Resolution with Internalized Calibration; the Ts-VI triangle model; and SSEB-Simplified Surface Energy Balance). They estimated ET in an oasis-desert region during a growing season where METRIC out-performed both the Ts-VI triangle and SSEB models. The Ts-VI triangle model tended to overestimate and the SSEB to underestimate at higher values of ET. In their paper, it is concluded that reasonable estimates of ET can be achieved by carefully selecting extreme pixels or edges, and validation is required when applying remote sensing-based models, especially the contextual methods.

Irmak et al. (2011) estimated the land surface ET at Nebraska. Fractional ET (ETrF) maps generated by METRIC were lower early in the growing season for agricultural crops and gradually increased as the NDVI of crops increased which results water to transpire more towards the midseason. Comparing the results with the Bowen ratio energy balance system showed that METRIC performed well at the field scale for predicting ET from a cornfield and if calibrated properly, the model could be used to estimate water use in managed ecosystems in sub-humid climates at a large scale.

Madugundu et al. (2017) estimated daily ET by the application of the METRIC model using Landsat 8 images in an irrigated field of Saudi Arabia. The authors compared the estimated ET against the EC flux tower ET and found that the METRIC estimated ET was quite close with the observed ET, with an RMSE of 0.13 mm/hr and 4.15 mm/day (instantaneous ET and daily ET). The authors also concluded that the METRIC performance was better in fully canopy areas than in the partial canopy areas.

Reyes-González et al. (2017) estimated ET from the satellite-based METRIC model and compared the results with in-situ Atmometer readings in South Dakota and found that the ET estimated by METRIC model showed a good relationship with the ET estimated by Atmometer with an RMSE of 0.65 mm/day. The authors concluded that the results were useful in the field of irrigation water management at local and regional scales.

Numata et al. (2017) estimated daily, monthly and, seasonal ET on Amazonian forests using Landsat based METRIC model. The results obtained were compared with flux tower data which showed a good correlation of $r^2 > 0.70$ and % MAE <15% in overall seasons. The authors concluded that the METRIC model is applicable for ET estimations despite the challenges in selecting cold anchor pixels in dry conditions, which is used during METRIC calibration. He et al. (2017) estimated ET in an Almond orchard field of California using the METRIC model, where the author found the obtained ET was close with the field observation with an RMSE of 0.11/hr in all seasons and 0.09 mm/hr in growing seasons.

The international panel on climate change (IPCC) declared that the potential ET rate rises with air temperature (IPCC, 2007). Recent studies during the past decades have shown that climate change and climate variability have caused changes in environmental controls on ET leading to widespread change in ET (Miralles et al., 2014; Zhang et al., 2016b). Vegetation has a major influence on ET besides these environmental controls. Different studies have been done on the influence of vegetation on ET and among them, some show that in the growing season, inter-annual change in vegetation activity predominantly control inter-annual changes in ET (Lawrence et al., 2011; Suzuki et al., 2007) while some other studies show that the influence of vegetation is affecting spatiotemporal patterns and trends in ET (Wang et al., 2010; Zhang et al., 2015b).

In summary, several works have been done using remote sensing models for ET estimation in different regions all around the world. In general, these studies have shown that remote sensing is capable to estimate ET in good agreement with ground-based ET estimation. Next, we discuss some of the earlier works that have focused on ET estimation and general analysis of ET distribution for regions in Nepal.

## 1.5 ET estimation for regions in Nepal

One of the first works concerning ET measurement and/or analysis in Nepal was done by authors in Lambert and Chitrakar (1989). This study in Nepal concerning the variation of potential evapotranspiration (PET) for each month regressed against elevation (which ranged from 100 to 3,700 m) gave quite consistent, good correlations. The seasonal variation of PET and its elevation gradient showed systematic and reliable results with the seasonal and elevation patterns of related climate parameters. In another study, the variation of the most important climatic variables during 2000–2016 along the North-South transect across Jharkhand, Bihar and Eastern Nepal, showed the increasing trend in the case of ET and a wider stretch of net radiation values due to its highly undulating topography. Mahto and Pandey (2018) concluded that due to the increasing trend of surface ET good agricultural land transfer into fallow land in the future which will be a serious issue for both farmers and local livelihoods of that region.

In western Nepal, the air temperature has steadily warmed by almost 1°C since 1990 which acts to enhance ET, which exacerbates any drought that has already been induced by any precipitation decline (Wang et al., 2013). Likewise, due to global warming, the rate of ET is increased in the context of Nepal also (Chaudhary and Aryal, 2009). In a study for the period during 1981-2007 over central Nepal, it was observed that the temperature in all the region had an increasing trend in the past, indicating a great consequence for drought conditions along with an increasing water demand as a result of increasing ET in this region (Dahal et al., 2016). The ET increases due to an increase in temperature in pre-monsoon and monsoon season without adequate rainfall leading to a soil-moisture deficit and limiting tree growth (Fritts, 2012).

In a study of ET from the natural and planted forest in the middle mountains of Nepal by Baral (2012), it was concluded that the ET from planted pine forest is large than the natural forest or degraded land which results in the drying of water resource in the middle mountain. Ghimire et al. (2014) studied the transpiration and canopy conductance of two contrasting forest types in the Lesser Himalaya of central Nepal. This study concluded that the average daily transpiration rate and canopy conductance were higher in the pine forest than in the natural forest whereas transpiration rates in both forests increase in the dry season indicating that the roots must have access to deeper soil layers and the weathered portion of the geological substrate. The higher ET obtained for the planted coniferous forest relative to degraded pasture in the study area observed by Baral (2012) and the higher transpiration during the dry season observed by Ghimire et al. (2014) support the decline in dry season base flows which leads to the large scale reforestation of degraded grass and scrubland in the Middle Mountains of Central Nepal.

There have also been earlier studies on the use of remote sensing for ET estimation in the region of Nepal. For example, Amatya et al. (2015) investigated the trend of land surface heat flux (energy balance components relevant for the ET estimation) for the southern slope of the central Himalayas in Nepal. MODIS dataset using the SEBS model was used for the analysis. The authors pursued trend analysis from the year 2003 to 2013. The land surface heat flux in the Central Himalaya was found to be increasing over the years as a result of decreasing precipitation. In another study, Amatya et al. (2016) used MODIS data to estimate land surface heat fluxes over Nepal using the Topographically Enhanced Surface Energy Balance System (TESEBS) model. The authors concluded that land surface heat flux varies according to seasons, with differences seen for pre-monsoon, monsoon, post-monsoon and winter seasons. Both earlier studies analyzing the regions of Nepal used MODIS data. These results obtained showed the application of remote sensing for estimating spatial heat fluxes patterns over a very large region, something not scalable for ground-based field measurement approach. However, these results were based on satellite imagery with low spatial resolutions (5 Km) and thus were not able to estimate finer spatial variations. In a topographic diversified region like Nepal, high-resolution satellite imagery is needed to better reflect the variations of ET and other climatic conditions.

## 1.6 Research questions

As Nepal is a topographic complex region, extending from plain flat lands to high hill and mountains within a small geographic region, remote sensing provides a viable alternative for ET estimation on a large scale. However only a few researches have been conducted on ET estimation for Nepal using remote sensing, and none of these have investigated the use of high-resolution satellite imagery such as those afforded by Landsat 8. Further, given the topographical variations in Nepal, especially with large ranges of elevation (57 m - 8844 m) in a small land-span from north to south (193 km on average), our study area allows generating insights on the elevation-wise variation of ET. These insights could shed further light on how different topographies have an impact on ET and what model adaptations are required, if any, to work across the ranges of topography. Thus, our research in this thesis is guided by following two broad questions:

- Can high-resolution satellite imagery provide a viable alternative to large scale ET estimation in Nepal?
- How does ET vary with elevation for a region like Nepal, across different seasons/months, with highly differing topographical features at these elevations?

## 1.7 Objective of the research

The objective of our research is summarized as:

- To utilize the high-resolution satellite imagery of Landsat 8 and compute hourly and daily ET for Nepal
- To analyze the elevation-wise variation, as seen in different months across the seasons, for a topographic diversified region such as Nepal
- To implement and validate the METRIC model that needs only a few ground-based measurements for the ET estimation, towards the objective of obtaining ET and analyzing elevation-wise variations of ET

Previous studies have shown the feasibility of using remote sensing-based methods for ET estimation over a large region. In this study, we present our analysis on ET estimation using Landsat 8 data, which provides a very high spatial resolution of 30m. Our study area is Nepal which represents a region with high topographical variations. METRIC model is used for computing ET in our work. METRIC model has been shown to produce high-quality accurate maps for ET from remotely sensed data in the visible, near-infrared and thermal infrared spectral regions along with ground-based wind speed and near-surface dew point temperature (Allen et al., 2007). Surface slope, aspect and, temperature lapsing is considered within the METRIC model, making the model applicable for both flat and mountainous terrain as is the case for our selected region. In METRIC modeling, satellite-based crop classification is not required, as is required in other surface energy balance models, and the calibration of the model is done via reference ET which improves accuracy. Each satellite image is auto-calibrated using weather station-based data in the METRIC modeling. To the best of our knowledge, our work is the first that is investigating

ET estimation for a topographically diverse region like Nepal using Landsat data and METRIC model, thus producing a high-resolution ET estimation.

## 1.8 Outline of the thesis

This thesis is organized into five different Chapters. In this first Chapter, a general introduction has been given. Different relevant literature on ET estimation and the objectives of the current research is also discussed in the same Chapter. In Chapter 2, we describe the study area, datasets, and data processing methods used for the analysis in current work. All the required methodology for the computation of different parameters such as heat fluxes, vegetation indices, surface temperature, surface albedo are presented. Also, our model setup is discussed in this chapter. Validation of the METRIC model and associated  results obtained from the  model are described in Chapter 3. Elevation wise variation of ET along with $T_s$ and NDVI is discussed in Chapter 4. The conclusion based on robust findings of current study are presented in  in Chapter 5.

# Chapter 2 Materials and Methods

## 2.1 Study area

The study area considered for our work was Nepal, a small mountainous and landlocked country between India and China. The study area is located between 26.35°N to 30.45°N and 80.06°E to 88.20°E covering an area of 147,181 km$^2$. Our considered region extends from 57 m above sea level (m asl) up to 8844 m asl. The study area with the location of the EC (eddy covariance) station used for our work is shown in **Figure 4**. The elevation in the region increases from south to the north in general and thus the temperature decreases from south to the north as the temperature is strongly influenced by elevation. The region, and Nepal in general, extends parallel to the Himalayan range in the north of the country. Seasons in our considered region are divided based on monsoon circulation pattern into four seasons: pre-monsoon (March, April, May), monsoon (June, July, August, September), post-monsoon (October, November) and winter (December, January, February) (Nayava, 1980). There is a large variety of land types with elevation in this region.

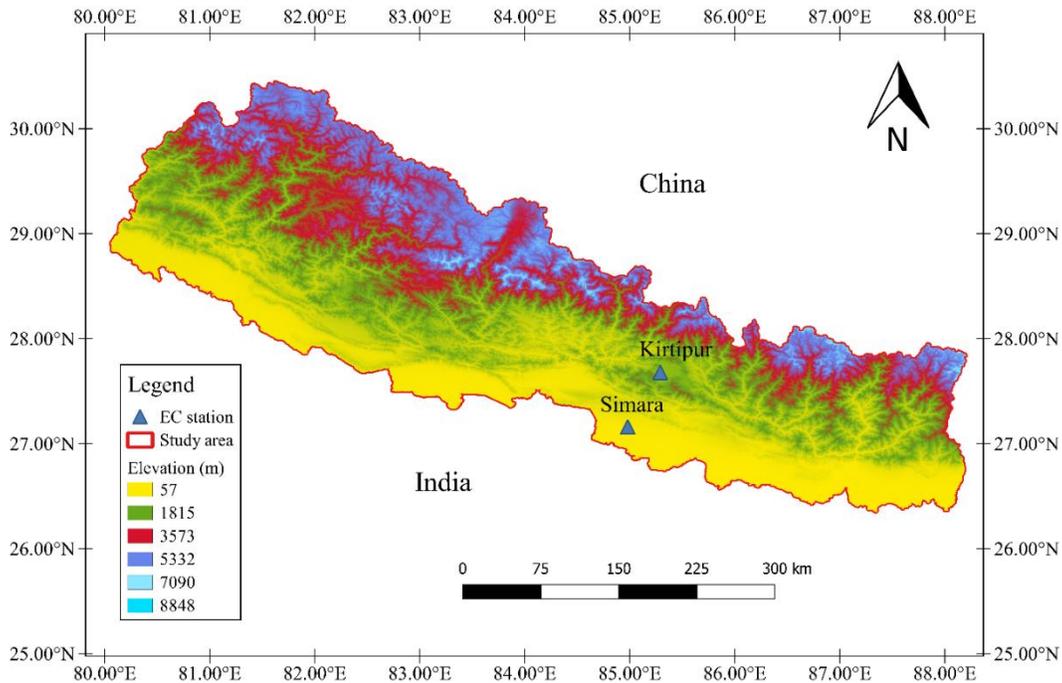

**Figure 4** Study area with the location of EC station considered for our ET calculations. The color map shows different elevation distribution.

Approximately 83% of the area is composed of mountains and hills which includes diversified geomorphic types like flat terraces, valleys, hills, snow-peaked mountains. Such diversity generates varied climatic zones within the small areal extent, stretching from sub-tropical climate in the southern plains to tundra climate in the high mountains. The climate of Nepal is dominated by the Indian Ocean monsoon climate, particularly summer monsoon-borne precipitation is the key feature of Nepal's climate. The region on an average receives about 1800 mm of annual precipitation, of which 79.8% is during the summer monsoon (DHM, 2017). Monsoon rainfall decreases from south-east to north-west owing to the distance from the Bay of Bengal and orography. In contrast, winter is relatively dry. Though Mediterranean westerlies bring winter rainfall, it is light and decreases from west to east. The southern plains have a sub-tropical climate with hot summer and mild winter while high mountains in the north have a tundra climate with subfreezing temperatures throughout the year.

## 2.2 Data

The selected study area for this work spans both flat and mountainous regions. Due to the unavailability of the meteorological data in the higher elevation areas, and to get an estimation at high-resolution, in this work, remote sensing data has been used. Remote sensing data used in this works are from Landsat 8, and Landsat 5. From Landsat 8 both the sensors (Operational Land Imager (OLI) and Thermal Infrared Sensor (TIRS)) data are used. From Landsat 5, data from the sensor Thematic Mapper is used. Advanced Space-borne Thermal Emission Reflectance Radiometer Global Digital Elevation Model (ASTER-GDEM) is also used in this work. Not only remote sensing data but also meteorological data from the EC flux tower located on Kirtipur and Simara is used in this work. All the data used for our work i.e. both remote sensing data, as well as meteorological data, will be discussed in further detail in Section 2.2.1 and Section 2.2.2.

### 2.2.1 Remote sensing data

#### 2.2.1.1 Landsat 8

Over the past few decades, Landsat has been one of the primary operational earth observation satellites. Due to high spatial resolution, Landsat has been widely utilizing for both research and non-research purposes. Recently, NASA/USGS launched Landsat 8 on February 11, 2013 with two sensors onboard, Operational Land Imager (OLI) Thermal Infrared Sensor (TIRS) which collects data nearly coincidentally. OLI contains nine spectral bands, seven from Thematic Mapper (TM) and Enhanced Thematic Mapper Plus (ETM+) sensors from earlier Landsat satellite and two new spectral bands a deep blue coastal/aerosol band and a shortwave-infrared cirrus band. TIRS consists of 2 different Thermal infrared bands (TIRS 1 and TIRS 2). Landsat 8 has a temporal repeat cycle of 16 days and is referenced to the WRS-2 system.

Landsat 8 satellite imagery with a spatial resolution of 30 m for the year 2016 with less than 10% cloud cover is used in this research. The required images for the study area are obtained from the United States Geological Survey (USGS) Earth Explorer site (http://earthexplorer.usgs.gov/). To cover the entire study area, 13 Landsat-8 tiles are required with different paths and rows shown in **Figure 5**. Different bands from OLI and TIRS sensors used in this work with their wavelength are shown in **Table 2**.

**Table 2** Different bands used from Landsat 8 with their wavelengths

| Bands | Wavelength (micrometers) |
|---|---|
| Band 2 – Blue | 0.452 - 0.512 |
| Band 4 – Red | 0.636 - 0.673 |
| Band 5 - Near Infrared (NIR) | 0.851 - 0.879 |
| Band 6 - Shortwave Infrared (SWIR) 1 | 1.566 - 1.651 |
| Band 7 - Shortwave Infrared (SWIR) 2 | 2.107 - 2.294 |
| Band 10 - Thermal Infrared (TIRS) 1 | 10.60 - 11.19 |

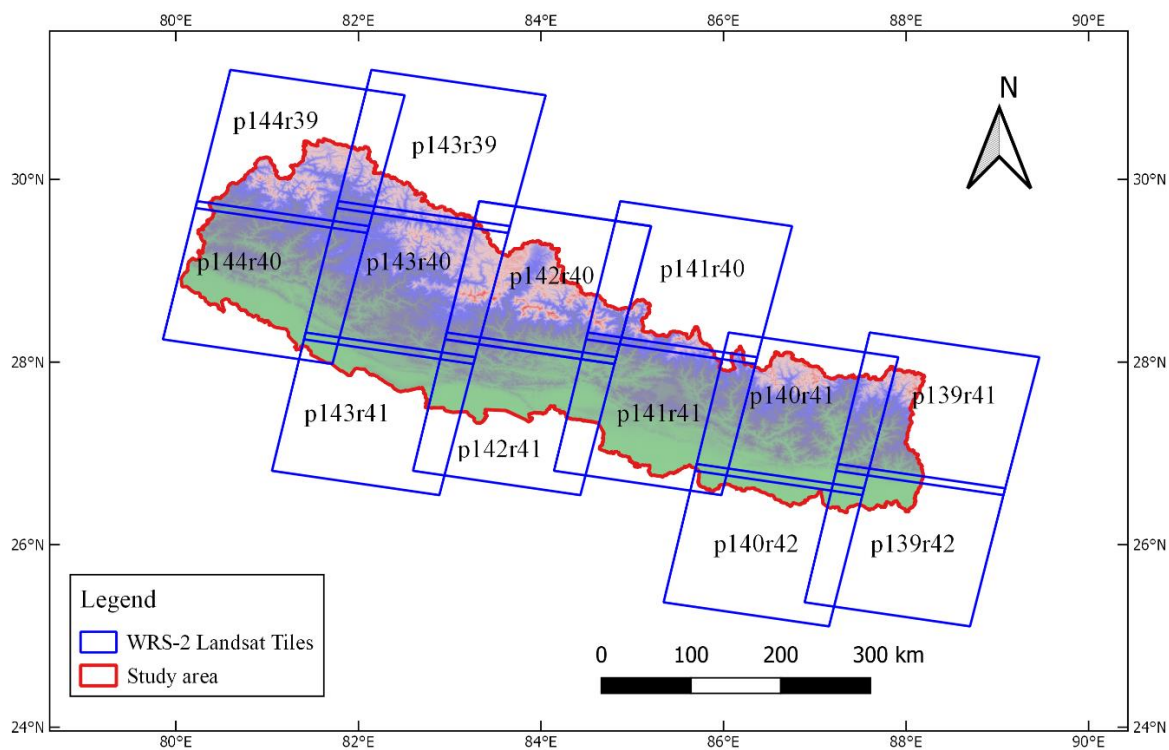

**Figure 5** Landsat 8 tiles covering the study area (Nepal).

We processed 26 Landsat 8 tiles to cover the entire study area for two months (March and October) during 2016. Moreover, we processed another 8 different Landsat 8 tiles to observe the elevation wise variation for 6 different months (March, April, May, June, October, and November). Since the Landsat 8 has a temporal repeat cycle of 16 days, the satellite doesn't overpass at the same date for all path and row of the study area. So, we considered the closest date across the month for different paths and rows. Path, row and, date of satellite overpass of our study area is shown in **Table 3**.

**Table 3** Path, row, and date of satellite overpass of Landsat 8 over the entire study area

| Path | Row | Date of satellite overpass |
|------|-----|----------------------------|
| 144 | 39 | 20 March, 30 October 2016 |
| 144 | 40 | 20 March, 30 October 2016 |
| 143 | 39 | 29 March, 23 October 2016 |
| 143 | 40 | 29 March, 23 October 2016 |
| 143 | 41 | 29 March, 23 October 2016 |

| 142 | 40 | 22 March, 01 November 2016 |
| 142 | 41 | 22 March, 01 November 2016 |
| 141 | 40 | 15 March, 16 April, 18 May, 3 June, 25 October, 10 November 2016 |
| 141 | 41 | 15 March, 16 April, 18 May, 3 June, 25 October, 10 November 2016 |
| 140 | 41 | 24 March, 18 October 2016 |
| 140 | 42 | 24 March, 18 October 2016 |
| 139 | 41 | 17 March, 27 October 2016 |
| 139 | 42 | 17 March, 27 October 2016 |

### 2.2.1.2 ASTER-GDEM

The Digital Elevation Model (DEM) used in the research is derived from the Advanced Space-borne Thermal Emission Reflectance Radiometer (ASTER). ASTER global digital elevation model (GDEM) are arranged into tiles, each covering one degree of latitude and one degree of longitude, named according to their southwestern corners. ASTER-GDEM in the different resolution of 1 Arc–Second (30 m) is downloaded from the USGS website (https://earthexplorer.usgs.gov/ ). This is necessary for use in the geographical information system to obtain elevation, slope, and aspects of the study area. During METRIC processing, this DEM was used to adjust surface temperatures for lapse effects caused by elevation variation, radiation flux as a function of elevation, and to calculate momentum roughness and wind speed per pixel.

### 2.2.1.3 Land Use Land Cover map

The land cover map for the study area used in this research is obtained from Uddin et al. (2015) as shown in **Figure 6**. For land cover mapping, 11 scenes (185 by 185 km each) of Landsat TM satellite images of 30 m (2010) data was used. According to this land cover map, forest and natural vegetation are the dominant land cover in lower elevation whereas higher elevation constitutes of barren area and snow/glaciers as major land cover. Land use map was used for developing spatial variations in surface roughness and for the condition selection for 'hot' and 'cold' pixel during the METRIC model based processing.

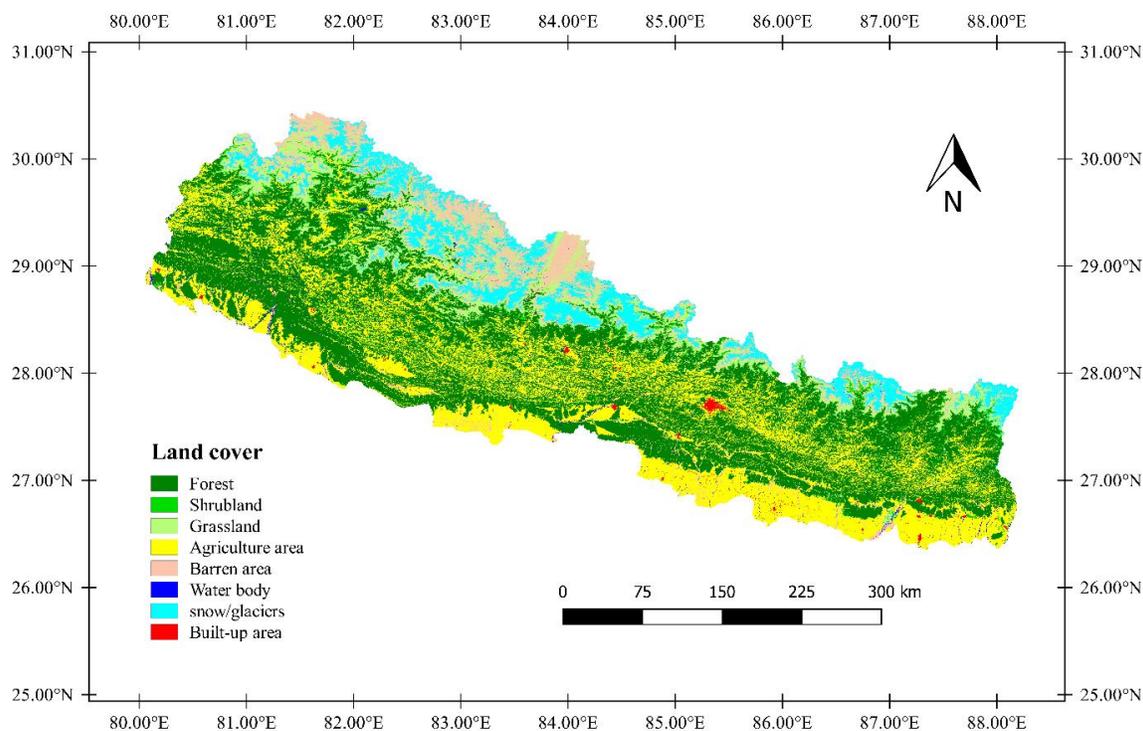

**Figure 6** Land use and land cover for the study area. The color map shows different land type distributions.

### 2.2.2 Meteorological data

We used the data from EC flux tower (established by cooperation between Institute of Tibetan Plateau Research, Chinese Academy of Sciences (ITPCAS) and Tribhuvan University) at two different stations (Kirtipur and Simara), as ground station meteorological data. Different meteorological variables at these stations are measured by the sensors, installed in a 40 m tall planetary boundary layer tower. Relative humidity, air temperature, wind speed, and wind direction are measured above the ground level by the sensors installed at 40 m, 6 m, and 3 m. Soil heat flux is measured at a depth of 5cm. Incoming and outgoing shortwave and longwave radiation were measured by the net radiometer installed at 1.5 m above the ground level. The details about the data processing can be found in Joshi et al. (2020).

Wind speed and near-surface vapor pressure from Kirtipur station (used for model calibration) are employed for the computation of reference sensible heat flux. Besides these, air temperature, soil heat flux, net radiation, hourly wind speed at 2 m height, and the saturation and actual vapor pressure (kPa) measured by the station are used for the calculation of the reference ET in model calibration. Similar meteorological parameters from Simara station were used to compute the ET for validation. For getting relevant information for a study area, e.g. identifying the estimated ET for the station area, an average of the estimations available within and surrounding an area (if no estimations are available in the region, e.g. due to cloud) can be used. Elevation, latitude, and longitude of these two stations are shown in **Table 4**. Kirtipur station is situated in the grassland area and the average canopy height is 0.5 m while the Simara station is mainly covered with grasses.

**Table 4** Location of eddy covariance flux tower site measurement of two different station

| Location | Elevation (m asl) | Latitude | Longitude |
|----------|-------------------|----------|-----------|
| Simara | 137 | 27.16˚N | 84.98 ˚E |
| Kirtipur | 1318 | 27.68 ˚N | 85.29 ˚E |

## 2.3 Landsat preprocessing

Landsat 8 Collection 1 level-1 data, which is used for our analysis, is in the form of a quantized digital number (DN) which is converted to at-sensor radiance and top of atmosphere (TOA) reflectance with a correction for solar zenith angle. The detail information about Landsat 8 is available at Landsat 8 Data Users Handbook, 2016 (https://Landsat.usgs.gov/Landsat-8-l8-data-users-handbook).

The DN is converted to at-sensor radiance as:

$$L_\lambda = ML \times Q_{cal} + AL \tag{2.1}$$

where $L_\lambda$ is TOA radiance without correction for the solar angle, ML is Band-specific multiplicative rescaling factor, AL is Band-specific additive rescaling factor, $Q_{cal}$ is DN

TIRS band data is converted to brightness temperature using Landsat-8 product guide by USGS, 2013 (http://Landsat.usgs.gov/Landsat8) as:

$$TB = \frac{K_2}{\ln\left(\frac{K_1}{L_\lambda} + 1\right)} - 273.15 \tag{2.2}$$

where $K_1$ and $K_2$ are band-specific thermal conversion constant.

The quantized pixel values are converted to TOA reflectance without correction for solar angle by using the equation as

$$\rho'_\lambda = M_\rho \times Q_{cal} + A_\rho \qquad (2.3)$$

where $\rho'_\lambda$ is TOA reflectance without correction for the solar angle, $M\rho$ is band-specific multiplicative rescaling factor, $A\rho$ is band-specific additive rescaling factor.

The TOA reflectance with a correction for solar zenith angle is given as:

$$\rho_\lambda = \frac{\rho'_\lambda}{\cos\theta_{SZA}} \qquad (2.4)$$

where $\theta_{SZA}$ is the Solar Zenith angle.

## 2.4 METRIC model

We used the METRIC model for ET estimation in our work using Landsat 8 data. METRIC is a satellite-based image-processing surface energy balance model for the estimation of regional evapotranspiration over complicated surfaces (Allen et al., 2007). METRIC has been extended from SEBAL developed by Bastiaanssen et al. (1998a) through integration with reference ET, which is computed using ground-based weather station data. In the METRIC model, ET is estimated as a residue from the energy balance equation as:

$$LE = R_n - G - H \qquad (2.5)$$

where $R_n$ is the net radiation, G is the soil heat flux, H is the sensible heat flux and LE is the latent heat flux consumed by ET. All the fluxes are expressed in watts per meter square ($Wm^{-2}$). We describe the procedure for computing different energy fluxes next.

### 2.4.1 Net radiation

$R_n$ is computed by subtracting all outgoing radiant fluxes from all incoming radiant fluxes (Brunsell and Gillies, 2002; Hipps, 1989) as:

$$R_n = R_{S\downarrow} - \alpha R_{S\downarrow} + R_{L\downarrow} - R_{L\uparrow} - (1 - \varepsilon_o) \times R_{L\downarrow} \qquad (2.6)$$

where $R_{s\downarrow}$ is incoming broad-band short-wave radiation, $R_{L\downarrow}$ is incoming longwave radiation, $R_{L\uparrow}$ is outgoing longwave radiation, $\varepsilon_o$ is broadband surface thermal emissivity, and $\alpha$ is surface albedo.

$R_{s\downarrow}$ is the principle source of energy for ET and is calculated under clear sky conditions as

$$R_{s\downarrow} = \frac{G_{sc}\,cos\theta_{rel}\,\tau_{sw}}{d^2} \qquad (2.7)$$

where $G_{sc}$ is solar constant ($1367 Wm^{-2}$), $\theta_{rel}$ is solar incidence angle, $\tau_{sw}$ is broad-band atmospheric transmissivity, $d^2$ is square of relative Earth-Sun distance. $R_{s\downarrow}$ for a mountainous region like Nepal must be computed for each pixel of the scene considering slope and aspect.

$d^2$, the function of the day of the year (DOY) is calculated as in Duffie and Beckman (2013):

$$d^2 = \frac{1}{1 + 0.033 \cos\left(\frac{DOY2\pi}{365}\right)} \qquad (2.8)$$

$\theta_{rel}$ for each pixel is computed as (Allen et al., 2007):

$$\begin{aligned}
\cos\theta_{rel} &= \sin(\delta)\sin(\phi)\cos(s) - \sin(\delta)\cos(\phi)\sin(s)\cos(\gamma) \\
&\quad + \cos(\delta)\cos(\phi)\cos(s)\cos(\omega) \\
&\quad + \cos(\delta)\sin(\phi)\sin(s)\cos(\gamma)\cos(\omega) \\
&\quad + \cos(\delta)\sin(\gamma)\sin(s)\sin(\omega)
\end{aligned} \qquad (2.9)$$

where $\delta$ is the declination of the Earth, $\phi$ is the latitude of the pixel $s\ is$ surface slope (s=0 for horizontal, $s = \frac{\pi}{2}$ radians for vertical downward slope) $\gamma$ is surface aspect angle ($\gamma = 0$ for slopes oriented due south, $\gamma = -\frac{\pi}{2}$ radians for slopes oriented due east, $\gamma = \frac{\pi}{2}$ radians for slopes oriented due west, $\gamma = \pm\frac{\pi}{2}$ radians for slopes oriented due north) $\omega$ is the hour angle ($\omega = 0$ at solar noon, $\omega =$ negative in morning, $\omega =$ positive in afternoon).

Slope and aspect are computed from DEM and $\delta$ is computed as (Cooper, 1969):

$$\delta = -23.45° \times \cos\left(\frac{360}{365} \times (d + 10)\right) \qquad (2.10)$$

Broad-band atmospheric transmissivity ($\tau_{sw}$) is calculated from the general function of standardizations for calculating ET as ASCE-EWRI (2005)

$$\tau_{sw} = 0.35 + 0.627\exp\left[\frac{-0.00146P}{K_t\cos(\theta_{hor})} - 0.075\left(\frac{W}{\cos\theta_{hor}}\right)^{0.4}\right] \qquad (2.11)$$

where P is the atmospheric pressure (kPa), W is the amount of water present in the atmosphere (mm), $K_t$ is the air turbidity coefficient ($0 < K_t \leq 1.0$) and $\theta_{hor}$ is the solar zenith angle over a horizontal surface which is computed as

$$\cos\theta_{hor} = \sin(\delta)\sin(\phi) + \cos(\delta)\cos(\phi)\cos(\omega) \qquad (2.12)$$

P is calculated according to ASCE-EWRI (2005) as

$$P = 101.3\left(\frac{293 - 0.0065z}{293}\right)^{5.26} \qquad (2.13)$$

where, 293 is standard air temperature (K), and z is the elevation above sea level (m).

W is calculated using measured or estimated near-surface vapor pressure from a representative weather station (Garrison and Adler, 1990) as:

$$W = 0.14e_aP_{air} + 2.1 \qquad (2.14)$$

where $e_a$ is near-surface vapor pressure (kPa).

**Surface albedo**

Albedo comprises the broadband reflectance over the entire solar spectrum. Liang (2001) estimated broadband surface albedo from Landsat 5 using TOA reflectance bands and later Naegeli et al. (2017) calculated albedo from Landsat 8 with the same procedure as estimated by Liang (2001) and is given as:

$$\alpha = 0.356b_2 + 0.130b_4 + 0.373b_5 + 0.085b_6 + 0.072b_7 - 0.0018 \qquad (2.15)$$

where, $b_2$, $b_4$, $b_5$, $b_6$, $b_7$ are the TOA reflectance for band 2, 4, 5, 6, 7 from Landsat 8.

$R_{L\uparrow}$ emitted from the earth surface is driven by $T_s$ and surface emissivity and is computed using Stefan–Boltzmann equation:

$$R_{L\uparrow} = \varepsilon_o\sigma T_s^4 \qquad (2.16)$$

where, $\sigma$ is Stefan–Boltzmann constant ($5.67x10^{-8}Wm^{-2}$ $K^{-4}$), $T_s$ is the surface temperature (K), and $\varepsilon_o$ is broad-band surface emissivity.

Surface emissivity is calculated as in Tasumi (2003):

$$\varepsilon_o = 0.95 + 0.01\ LAI \quad for\ LAI \leq 3 \tag{2.17}$$

$$0.98\ for\ LAI > 3$$

**Leaf Area Index (LAI):**

LAI that characterizes the plant canopy is the ratio of one-sided green leaf area per unit ground surface area is computed from an empirical equation as given by Bastiaanssen (1998).

$$LAI = -\frac{\ln\left[\frac{(0.69 - SAVI)}{0.59}\right]}{0.91} \tag{2.18}$$

where SAVI is soil adjusted vegetation index based on the top of atmosphere reflectance bands 4 and 5 of Landsat 8 and is estimated as in Huete (1988):

$$SAVI = \frac{(1 + L)(band5 - band4)}{L + (band5 + band4)} \tag{2.19}$$

where L is the canopy background adjustment factor (0.5).

In METRIC, LAI is set to 6 when SAVI > 0.678 and LAI is set to zero when SAVI < 0.1.

**Land surface temperature (LST):**

LST is one of the important parameters that have direct effects on the variation of ET. It is computed using an empirical equation (Stathopoulou and Cartalis, 2007) as:

$$T_s = \frac{TB}{\left[1 + \left\{\left(\frac{\lambda TB}{\varrho}\right)\ln\varepsilon_\lambda\right\}\right]} \tag{2.20}$$

where $T_s$ is LST in Celsius (°C), λ is the average wavelength of band x, $\varrho = h * \frac{c}{\sigma} \sigma$ is the Boltzmann constant (1.38 x 10⁻²³ J/K), h is Plank's constant (6.626 x 10⁻³⁴), c is the velocity of light (3 x 10⁸ m/s) and ελ is emissivity computed as in Sobrino et al. (2004).

**Land surface emissivity (LSE):**

LSE largely dependent on the surface roughness and nature of vegetation cover and is calculated for the estimation of LST which is determined according to Sobrino et al. (2004):

$$\varepsilon_\lambda = \varepsilon_v \times P_v + \varepsilon_s (1 - P_v) + C \qquad (2.21)$$

where $\varepsilon_v$ is the emissivity for vegetation, $\varepsilon_s$ is the emissivity for soil, $P_v$ is the proportion of vegetation and C is the surface roughness ($C = 0$ for homogenous and flat surfaces) and taken as a constant value of 0.005 (Sobrino and Raissouni, 2000).

**Fractional vegetation (Pᵥ):**

Fractional vegetation coverage is obtained according to Carlson and Ripley (1997) as:

$$P_v = \left( \frac{\text{NDVI} - NDVI_{min}}{NDVI_{max} - NDVI_{min}} \right)^2 \qquad (2.22)$$

Where NDVIₘᵢₙ and NDVIₘₐₓ is the NDVI value for bare soil and full vegetation coverage.

**Normalized difference vegetation index (NDVI):**

Vegetation indices are necessary for the identification of the fully vegetative pixel, water, and dry bare soils. In METRIC, NDVI, LAI, and SAVI are generally calculated from TOA reflectance of the red and NIR bands. The land surface condition is characterized by NDVI and is computed using reflectance for near-infrared and red bands. For Landsat 8, band 5 is near-infrared (NIR) and band 4 is red (R).

$$NDVI = \frac{NIR - R}{NIR + R} \qquad (2.23)$$

Rₗ↓ is downward thermal radiation which is computed using Stefan–Boltzmann equation

$$R_{L\downarrow} = \varepsilon_a \sigma T_s^4 \qquad (2.24)$$

where $\varepsilon_a$ is effective atmospheric emissivity (dimensionless)

$\varepsilon_a$ is computed as (Allen et al., 2000; Bastiaanssen, 1995):

$$\varepsilon_a = 0.85(-\ln \tau_{sw})^{0.09} \tag{2.25}$$

where $\tau_{sw}$ is broad-band atmospheric transmissivity.

### 2.4.2 Soil heat flux (G)

Soil heat flux is the rate of thermal energy transfer through a unit area of soil over time due to conduction. In METRIC, G is calculated using an empirical equation developed by Tasumi (2003):

$$\frac{G}{R_n} = 0.05 + 0.18e^{-0.521\,LAI} \quad (LAI \geq 0.5) \tag{2.26}$$

$$\frac{G}{R_n} = \frac{1.80(T_s - 273.15)}{R_n} + 0.084 \quad (LAI < 0.5) \tag{2.27}$$

where $T_s$ is the surface temperature (K), LAI is the leaf area index.

### 2.4.3 Sensible heat flux (H)

H is computed as a function of aerodynamics resistance as:

$$H = \rho_{air}\, C_p \frac{dT}{r_{ah}} \tag{2.28}$$

where $\rho_{air}$ is air density (kg m$^{-3}$), $C_p$ is the specific heat of air at constant pressure (1004 Jkg$^{-1}$K$^{-1}$), dT is near-surface to the air-temperature difference between two heights ($z_1$ and $z_2$) in a near-surface blended area (K) $r_{ah}$ is aerodynamic resistance (s m$^{-1}$) between two near-surface heights.

$\rho_{air}$ should be computed pixel-wise and is calculated as Allen et al. (1998):

$$\rho_{air} = \frac{1000P}{1.01(T_s - dT)R} \tag{2.29}$$

where P is the mean atmospheric pressure for the pixel elevation (kPa), R is the specific gas constant (287 J kg$^{-1}$ K$^{-1}$)

For the calculation of H, there are two unknowns variable dT and $r_{ah}$ which need to be computed. dT is near-surface to the air-temperature difference between two heights ($z_1$ and $z_2$) in a near-surface blended area (K). dT is computed as done by Bastiaanssen (1995) as a linear function of land surface temperature ($T_s$):

$$dT = a \times T_{s\,datum} + b \qquad (2.30)$$

where a and b are empirically determined constant for a satellite image and $T_{s\,datum}$ is a surface temperature adjusted for a common elevation data for each image pixel using DEM and customized lapse rate.

$r_{ah}$ is first computed assuming neutral stability as (Allen et al., 2007):

$$r_{ah} = \frac{\left(\ln\frac{z_2}{z_1}\right)}{u^* \times k} \qquad (2.31)$$

where $z_1$ and $z_2$ are heights above the zero-plane displacement of the vegetation where the endpoints of the dT are defined (m), $u^*$ is friction velocity ($ms^{-1}$) and k is von Karman's constant (0.41).

$u^*$ is computed from logarithmic wind law for neutral atmospheric condition as:

$$u^* = \frac{k u_{200}}{\ln\left(\frac{200}{z_{om}}\right)} \qquad (2.32)$$

where, $u_{200}$ is the wind speed at a blending height assumed to be 200 m ($ms^{-1}$), and $z_{om}$ is the momentum roughness length (m).

Momentum roughness length ($z_{om}$) in METRIC for each pixel is computed according to land use type or amount of vegetation. $z_{om}$ computed as a function of LAI by using the empirical equation as given by Tasumi (2003) for agricultural areas:

$$z_{om} = 0.018 \times LAI \qquad (2.33)$$

For non-agricultural areas, $z_{om}$ is known and can be assigned using the land use map. Here in this work $z_{om}$ is computed pixel-wise by considering the slope as well (Tasumi, 2003):

$$z_{om_{mtn}} = z_{om}\left(1 + \frac{\left(\frac{180}{\pi}\right)s - 5}{20}\right) \tag{2.34}$$

Where, $z_{om\_mtn}$ is the momentum roughness length for mountainous areas, and s is the surface slope computed from DEM.

The wind speed at a blending height at 200 m is computed as:

$$u_{200} = \frac{u_w \ln\left(\frac{200}{z_{omw}}\right)}{\ln\left(\frac{z_x}{z_{omw}}\right)} \tag{2.35}$$

where $u_w$ is the wind speed measured at a weather station at $z_x$ height above the surface, and $z_{omw}$ is the roughness length for the weather station.

$z_{omw}$ is computed by the empirical equation develop by Brutsaert (1982) as:

$$z_{omw} = 0.12 \times h \tag{2.36}$$

where h is the height of the canopy at the weather station.

For our study area $u_{200}$ is computed as (Tasumi, 2003):

$$u_{200_{mtn}} = u_{200} \times \overline{\omega} \tag{2.37}$$

$\overline{\omega}$ is the wind speed weighting coefficient computed as:

$$\overline{\omega} = 1 + 0.1\left(\frac{Elevation - Elevation_{station}}{1000}\right) \tag{2.38}$$

where Elevation is the elevation of the pixel (m) for the image and Elevation$_{station}$ is the elevation where the wind speed is measured.

**Determination of constants a and b in the dT parameter:**

dT is computed by rearranging equation (2.30) and then the constants a and b are computed considering 'hot' and 'cold' anchor pixels for each satellite image.

At hot pixel, $dT_{hot}$ is computed as:

$$dT_{hot} = \frac{H_{hot} r_{ah\,hot}}{\rho_{air\,hot} C_p} \tag{2.39}$$

where $r_{ah\,hot}$ is aerodynamics resistance at the hot pixel, $\rho_{air\,hot}$ is the density of air at hot pixel and $H_{hot}$ is computed as:

$$H_{hot} = (R_n - G)_{hot} - LE_{hot} \tag{2.40}$$

where, $LE_{hot}$ is the estimated LE at hot pixel, and $R_{n\,hot}$ and $G_{hot}$ are the derived net radiation and soil heat flux at hot pixel.

At cold pixel, $dT_{cold}$ is computed as:

$$dT_{cold} = \frac{H_{cold} r_{ah\,cold}}{\rho_{air\,cold} C_p} \tag{2.41}$$

And sensible heat flux at cold pixel ($H_{cold}$) is computed as:

$$H_{cold} = (R_n - G)_{cold} - LE_{cold} \tag{2.42}$$

Where, $r_{ah\,cold}$ is aerodynamics resistance at cold pixel, $\rho_{air\,cold}$ is density of air at cold pixel, $LE_{cold}$ is the estimated latent heat flux at cold pixel, and $R_{n\,cold}$ and $G_{cold}$ are the derived net radiation and soil heat flux at cold pixel.

The coefficient a and b are then computed from dT and the $T_{s\,datum}$ at hot and cold anchor pixels as:

$$a = \frac{dT_{hot} - dT_{cold}}{T_{s\,datum\,hot} - T_{s\,datum\,cold}} \tag{2.43}$$

$$b = \frac{[dT_{hot} - a]}{T_{s\,datum\,hot}} \tag{2.44}$$

Here $T_{s\,datum\,hot}$ and $T_{s\,datum\,cold}$ are the surface temperature adjusted for common elevation data for each image pixel using DEM and customized lapse rate at hot and cold pixel respectively.

### 2.4.4 Calculation of Evapotranspiration

ET at the time of satellite overpass for each pixel is computed as in Allen et al. (2007):

$$ET_{inst} = 3,600 \frac{LE}{\lambda \rho_w} \qquad (2.45)$$

where $ET_{inst}$ is instantaneous ET (mm/hr), 3,600 is a time conversion from seconds to hours, $\rho_w$ is the density of water (1000 kg m$^{-3}$) and $\lambda$ is the latent heat of vaporization (J kg$^{-1}$). With this instantaneous ET estimation, daily and monthly ET can be computed as in Allen et al. (2007)

After the computation of $ET_{inst}$, reference evapotranspiration fraction ($ET_rF$) is computed for each pixel from reference ET ($ET_r$) which is computed from the weather station data.

$$ET_rF = \frac{ET_{inst}}{ET_r} \qquad (2.46)$$

Where, $ET_{inst}$, $ET_r$ and, $ET_rF$ is in mm/hr.

$ET_r$ is computed from standardized ASCE Penman-Monteith method (Allen et al., 2005) as:

$$ET_r = \frac{0.408\Delta(R_n - G) + \gamma \dfrac{C_n}{T + 273} u_2(e_s - e_a)}{\Delta + \gamma(1 + C_d u_2)} \qquad (2.47)$$

Where, $ET_r$ is reference ET (mm/hr), $\Delta$ is the slope of the saturation vapor pressure versus temperature curve (kPa °C$^{-1}$), $\gamma$ is the psychrometric constant (kPa °C$^{-1}$), T is air temperature (°C), $C_n$ (K mm s$^3$ mg$^{-1}$ h$^{-1}$) and $C_d$ (s m$^{-1}$) are the coefficients that vary with time, $u_2$ is the hourly wind speed at 2 m height (m s$^{-1}$) and, $e_s$ and $e_a$ are the saturation and actual vapor pressure (kPa).

### 24-Hour Evapotranspiration ($ET_{24}$):

Daily ET could be more useful than the instantaneous ET derived from the satellite image. $ET_rF$ has been used with 24-hour $ET_r$ to estimate the daily ET for each image pixel as:

$$ET_{24} = ET_rF \times ET_{r-24} \qquad (2.48)$$

where $ET_{24}$ is the daily value of actual ET (mm d$^{-1}$), $ET_{r-24}$ is 24-hour $ET_r$ for the day of the image, calculated by summing hourly $ET_r$ values over the day of image.

Thus, with the described procedure, one can use Landsat 8 data and other auxiliary information like calibration data in order to estimate ET and other parameters. Next, we describe how the METRIC model was implemented for our work.

## 2.5 Input Parameters for the METRIC model

METRIC is a surface energy balance model that needs four primary inputs to compute ET. Both satellite-based data and meteorological data are required for the implementation of the METRIC model. Here, in our work we use band 2,4,5,6,7,10 form Landsat 8, a DEM from ASTER-GDEM, Land use land cover map from ICIMOD (Uddin et al., 2015), and the meteorological data from EC flux tower located at Kirtipur. Detail about these four different primary inputs, we already discussed in Section 2.2. The working formulas and methodology were discussed in Section 2.4. Specific primary inputs for the model and the different surface parameters obtained from these primary inputs are given in **Table 5**.

**Table 5** Inputs parameters for the METRIC model

| Primary inputs | Surface parameters retrieved from different primary inputs |
| --- | --- |
| Landsat 8 | • Land surface temperature (band 10 is used to compute $T_s$) <br> • Surface albedo (band 2, 4, 5, 6, and 7 is used to compute surface albedo) <br> • Surface emissivity (band 4 and band 5 is used to compute surface emissivity) <br> • NDVI, SAVI, LAI (band 4 and 5 are used to compute NDVI, SAVI, and LAI) |
| ASTER-GDEM | • Slope (computed using RichDem library) <br> • Aspect (computed using RichDem library) <br> • Elevation |
| Land use land cover map | • Momentum roughness length <br> • Condition selection for hot and cold pixel |
| Meteorological data | • Wind speed at 200 m height (used for the computation of the sensible heat flux)<br><br>• near-surface vapor pressure ($e_a$) (used for the computation of the amount of water present in the atmosphere)<br><br>• air temperature, soil heat flux, net radiation, hourly wind speed at 2 m height, and $e_s$ and $e_a$ (saturation and actual vapor pressure (kPa)) are used for the calculation of the reference ET |

## 2.6 METRIC model setup

We implemented the METRIC model in Python3. The complete model setup is shown in **Figure 7**. In the figure, we show how different data was processed to compute intermediate parameters (like heat fluxes, $T_s$, vegetation index, surface albedo) and final estimation of ET. In the energy balance equation required for the computation of the latent heat flux (and subsequently the ET), the main energy components should be calculated are the net radiation, the soil heat flux, and the sensible heat flux. The primary inputs to compute these energy components are derived from the Landsat 8 data, DEM, Land use map, and a local weather station data.

Landsat 8 data were downloaded from the USGS website (https://earthexplorer.usgs.gov) for our region of interest. The download consists of the Digital Number (DN) values for different bands in the Landsat 8 and associated metadata information. The metadata information was parsed into a python dictionary data-structure which can be easily queried for any required metadata information. Some of the information derived from the metadata that is used in our model setup are satellite overpass time, solar zenith angle, and the latitude and longitude of the scene. The time information is referenced in the GMT time-zone and therefore converted to the local time-zone.

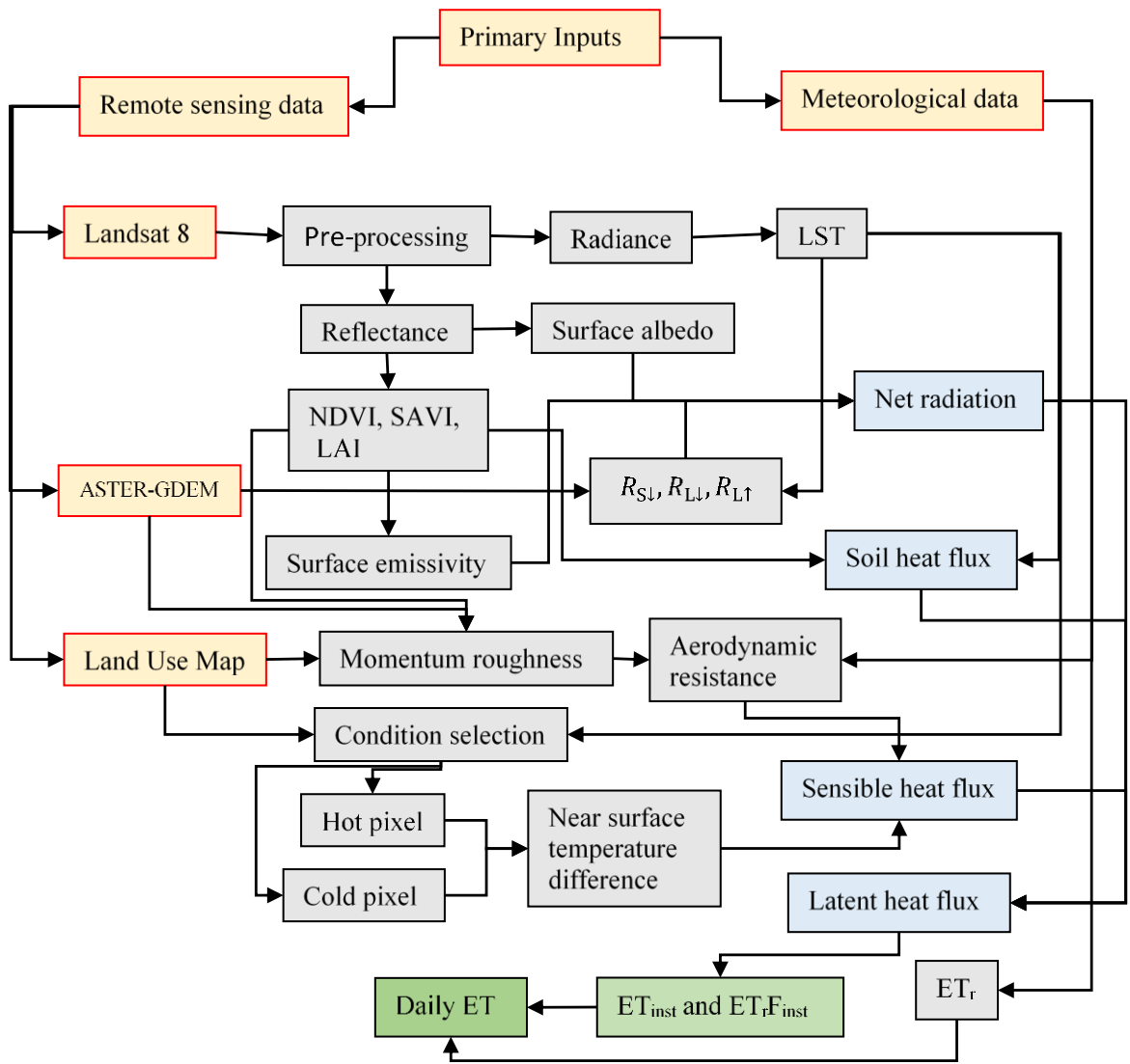

**Figure 7** Flow chart of complete METRIC model setup. Different colors as red, gray, blue and green indicate primary inputs, intermediate calculation, primary output, and final output respectively.

The weather station data is made available with a half-hourly recording of various weather parameters like air temperature, soil temperature, wind-speed, relative humidity, etc. Based on the time of the satellite overpass (after converting to the local time), the corresponding weather station data for that time is extracted in the model script. All weather station data points that are within 30 minutes from the satellite overpass time are extracted and their average is considered as the relevant weather information required in the METRIC model. Two data points are returned when processing to retrieve weather station data in our analysis. The following weather station information is used in our model setup: Relative Humidity, wind speed, vapor pressure deficit, air temperature, soil temperature, and LE as measured by the weather station. The latter has been extracted to provide a reference representative LE value for model validation.

Complete cloud-free Landsat 8 image is not available for 2016 for our study area. We used Landsat 8 images with less than 10% cloud cover. The presence of cloud has an impact on ET estimation. Importantly, the presence of cloud needs to be accounted for the selection of the hot and cold pixel which are part of the METRIC model. Selected hot and cold pixels need to come from cloud-free regions. We applied the Function of mask (Fmask) tool for the Landsat 8 image in Quantum geographic information system (QGIS) for cloud detection. The cloud mask from the detected cloud in the image is used to guide hot and cold pixel selection from the cloud-free area only. In our work, we selected the hot and cold pixels automatically by using the land use map and $T_s$. A hot pixel is taken from barren areas with high temperature and the cold pixel is selected from agricultural areas with low temperature, near the reference weather station (station used for calibration, not for validation).

The computation of variables required for the METRIC model was implemented within our python script. Net radiation computation requires all the incoming and outgoing radiation fluxed to be assessed. These have been discussed in Section 2.4.1. Several parameters need to be computed before these radiation fluxes can be obtained. These parameters are for example Land surface temperature, LAI, NDVI, fractional vegetation, land surface emissivity. The computation of Land Surface Temperature ($T_s$) will be described in further detail as an example to show how different parameters are computed within our model setup.

Computation of $T_s$ is based on the brightness temperature which in turn is computed from the Landsat image. The band 10 image is used for computing the brightness temperature (TB). The image is read using python's image read functionality and then pre-processed to calibrate based on the band's calibration parameter. These calibration parameters are retrieved from the metadata file associated with the image. Thanks to our processed metadata information available in a python dictionary, the calibration parameters can be easily accessed by referring to the band number alone. Once the band 10 image is calibrated, computing TB is straightforward after converting the calibrated image data to TOA radiance and using the known relationships that translate the TOA radiance to TB. Besides the TB, $T_s$ computation requires land surface emissivity information. This information is derived from the NDVI which is computed using the TOA reflectance computed from band 4 and band 5 image. In computing land surface emissivity, further assessment of NDVI for soil and vegetation is required. This can be obtained by mapping the computed NDVI to the land use map. In our setup currently, we just regard the maximum of the NDVI as coming from a vegetation area and the minimum from the soil area. The lowest 10 and highest 10 values are considered for this maximum and minimum computation (by taking their mean). This has been done to avoid any outliers being considered for the value of NDVI for vegetation or soil area.

After computing different parameters required for the energy fluxes, computing the net radiation fluxes and the soil heat flux is straightforward. In our model setup, we rely on various literature to set the value of the constants. For instance, we use the value of L to 0.1 as specified by Allen et al. (2007) to compute SAVI. The computation of sensible heat flux is a bit involved and thus is further elaborated here in the description of the model setup.

Two main parameters required for the sensible heat flux computation are dT (difference in temperature between two near land surface points) and $r_{ah}$ (aerodynamic resistance to heat transport). dT is computed by implementing a known linear regression model that translates land surface temperature (after elevation-based correction) to dT. The regression parameters are obtained by computing the sensible heat flux for the hot and cold pixels. In our current model setup, we compute the hot and the cold pixel automatically based on the land surface temperature and the land use map. The pixels from barren land which have the highest surface temperatures are taken as the hot pixel. Similarly, pixels corresponding to vegetation area as per the land use map and with the lowest surface temperatures are taken to be indicative of the cold pixels. We again take 10 pixels each as representing the hot and cold pixels to avoid any likely outliers when only one pixel is chosen. The known ET fraction for cold and hot pixels, as reported in the literature, is used to compute the sensible heat flux and correspondingly derive the dT for these extreme pixels. Once the dT and $T_s$ for a cold and hot pixel are available, it can be fit to the regression model to identify the final regression parameters to be used to map any $T_s$ to dT. Sensible heat flux value is computed using the computed value of dT. Sensible heat flux value together with the net radiation and soil heat flux are used to derive the latent heat flux and instantaneous ET.

It is to be noted that the model setup also uses a DEM which is required as input in the METRIC model, for example, to correct land surface temperature based on elevation (using a fixed lapse rate). The model setup currently reads the DEM computed from the ASTER. This DEM should be processed manually to align with the Landsat image being processed. This is currently done outside the main model setup, using the QGIS functionality. In our model setup, any discrepancy between the alignment of the Landsat image and the DEM, which might arise because the DEM were manually aligned, is corrected by image interpolation and re-sizing. This needs to be worked out further for accurate correspondence of elevation information to a given Landsat image pixel. Further, an accurate estimate of ET for a topographically diverse region like Nepal requires that many parameters like slope, aspect, momentum roughness, and wind speed be properly accounted for by the model. For example, to compute incoming solar radiation Rs↓, each pixel of the scene should be accounted for being a possible mountainous area. In our model implementation, the Solar incidence angle ($\theta_{rel}$) required for estimating $R_{s\downarrow}$ is computed for each pixel of the scene using slope and aspect also derived from the DEM. We have used the RichDEM library for computing slope and aspect. ET is also sensitive to aerodynamics which has direct effects on momentum roughness. For a complex terrain, direction of airflow and speed is difficult to measure but has direct impact on sensible heat flux (H) and Latent heat flux (LE) (Allen et al., 2013). Wind speed and momentum roughness length is thus computed in our model setup for each pixel by considering the elevation using DEM.

In summary, a basic model setup has been implemented that considers all the relevant parameters and computation for METRIC. We used this model implementation to compute ET for the region of Nepal which is described in the next Chapter.

# Chapter 3 Estimation of ET for Nepal using METRIC model

## 3.1 Background

As we discussed in earlier chapters, Nepal has large topographical variations and elevation ranges despite being only a small country. The range of elevation in Nepal is from 57 m to 8844 m in altitude. Large scale monitoring of weather parameters like ET has huge applications for a country like Nepal in terms of agriculture planning, climate and weather monitoring, and forecast, etc. The Himalayas which form the Northern boundary of Nepal is a major factor for climate regulation in the whole south-Asian region. Therefore, monitoring of ET in Nepal, closest to the Himalayas, also provides an insight into changing roles that Himalaya is playing for overall climatic regulations in the region. Despite the usefulness of large-scale monitoring of ET, it is difficult to use ground-based measurement stations for that purpose. Especially in the higher elevations, it is difficult to have ground-based weather stations due to inaccessibility and lack of supporting infrastructures. The cost factor associated with installing and maintaining numerous weather stations is also prohibitive. Therefore, for the area like Nepal, remote sensing-based technology could be highly effective to estimate ET. In Chapter 2 we introduced the METRIC model for mapping satellite imaging to ET estimations. We described the Landsat-8 based satellite images that we have used in this work for ET estimation, and all the processing steps involved in the METRIC model to get to ET estimation starting with the raw sensor images and some ground station data for calibration. We also described how we have implemented the METRIC model considering slope, aspect, and the temperature lapse rate as per the requirements to compute ET on a mountainous area more accurately. In this Chapter we present the validation of our model. Validation of our model was done to see whether the METRIC model-derived results are close enough to the observed results from the EC station located at Simara. It is to be noted that we have used the station data from Kathmandu for model calibration. Therefore, the use of Simara station for validation provides an independent validation setting. In this Chapter, we also present the ET estimation results for the entire region of Nepal.

## 3.2 Method

The metric model has been implemented as per the setup described in Chapter 2.4 and Chapter 2.5. The model estimated parameters such as ET was obtained for the region where the Simara station is located. This is then compared with the measurements obtained from the station data. Similarly, ET was estimated for the entire region of Nepal for two different months. We obtained the ET estimation for the month of March and October to get representations of two different seasons. Including more months is not straightforward as relatively cloud-free images need to be obtained for that corresponding months for all the tiles covering Nepal unless modeling is designed to be robust for the presence of cloud. The estimated ET for the whole region of Nepal is plotted for qualitative analysis. Similarly, we also compute simple statistics like average of ET, NDVI, $T_s$ over the whole region of Nepal and compare these statistics across the two months.

For calibration and validation of the model, two different flux tower station data are used. Kirtipur station data was used for the calibration purpose and Simara station was used for the validation purpose. As per the availability of the data, we validate surface temperature ($T_s$) and ET using root mean square error (RMSE) and mean bias error (MBE) as:

$$\text{RMSE} = \sqrt{\frac{\sum_{i=1}^{n}(\text{METRIC ET} - \text{fluxtower ET})^2}{n}} \tag{3.1}$$

$$\text{MBE} = \frac{\sum_{i=1}^{n}(\text{fluxtower ET} - \text{METRIC ET})}{n} \tag{3.2}$$

where n is the number of the sample taken.

## 3.3 Results

### 3.3.1 Validation of the METRIC derived $T_s$ and ET

We implemented the METRIC model to estimate ET using high-resolution Landsat 8 data for Nepal. To check the accuracy of the model, we first validated our METRIC model-derived results for different months with EC station located in Simara. As per the availability of the data from the EC station at Simara, we calculated RMSE and MBE for ET and $T_s$. The METRIC derived ET and $T_s$ were close to the flux tower ET and $T_s$ as shown in **Figure 8**. RMSE for hourly ET and daily

ET is obtained as 0.06 mm/hr and 1.24 mm/day while MBE for hourly and daily ET was 0.03 mm/hr and 0.29 mm/day respectively. RMSE and MBE for $T_s$ were obtained as 4.93 ˚C and -0.49 ˚C respectively. From the validation results, we can say that the METRIC model is applicable for the mountainous region like Nepal which is our study area.

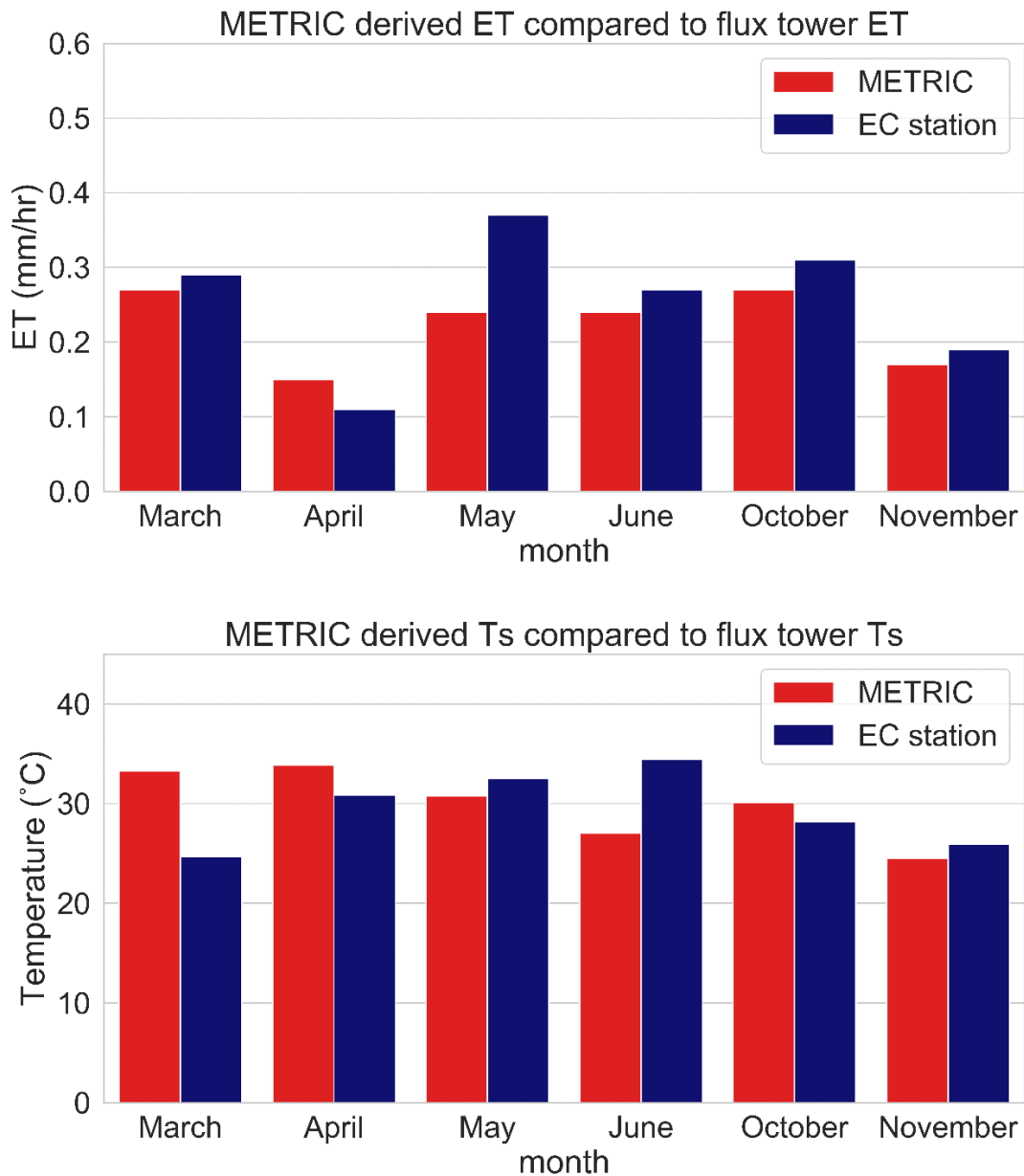

**Figure 8** Comparison of METRIC derived ET and $T_s$ with EC station ET and $T_s$ against different months.

### 3.3.2 Estimation of ET for the whole region of Nepal

We implemented the METRIC model and computed different parameters such as land surface temperature ($T_s$), vegetation indices (NDVI, SAVI, LAI), surface albedo, land surface heat fluxes ($R_n$, G, H, LE) and, finally ET for two different months March and October of the year 2016 for the entire Nepal region. This was done to see the elevation wise variation on ET and to compute hourly and daily ET. At first, we computed $T_s$ for March and October at the time of satellite overpass for Nepal. The average $T_s$ obtained from the METRIC model for March and October was obtained as 18.695 ˚C, and 19.112 ˚C. The overall $T_s$ for the entire Nepal region in the month of March and October as obtained by our model is shown in **Figure 9**. The variation in $T_s$ was seen from lower to a higher elevation (south to north). $T_s$ in October for the whole Nepal is seen to be high than in the March.

As a part of the intermediate calculation, we computed NDVI from band 4 and band 5 of Landsat 8 for March and October of the year 2016. The Average NDVI for March and October at the time of satellite overpass was obtained as 0.302 and 0.460. Minimum NDVI for March was -0.4 and maximum NDVI was 0.8 while for October minimum NDVI was obtained as -0.7 and maximum as 0.8. The values of NDVI over the entire region of Nepal for the month of March and October are shown in **Figure 10**. NDVI is seen to be high in October as compared to March.

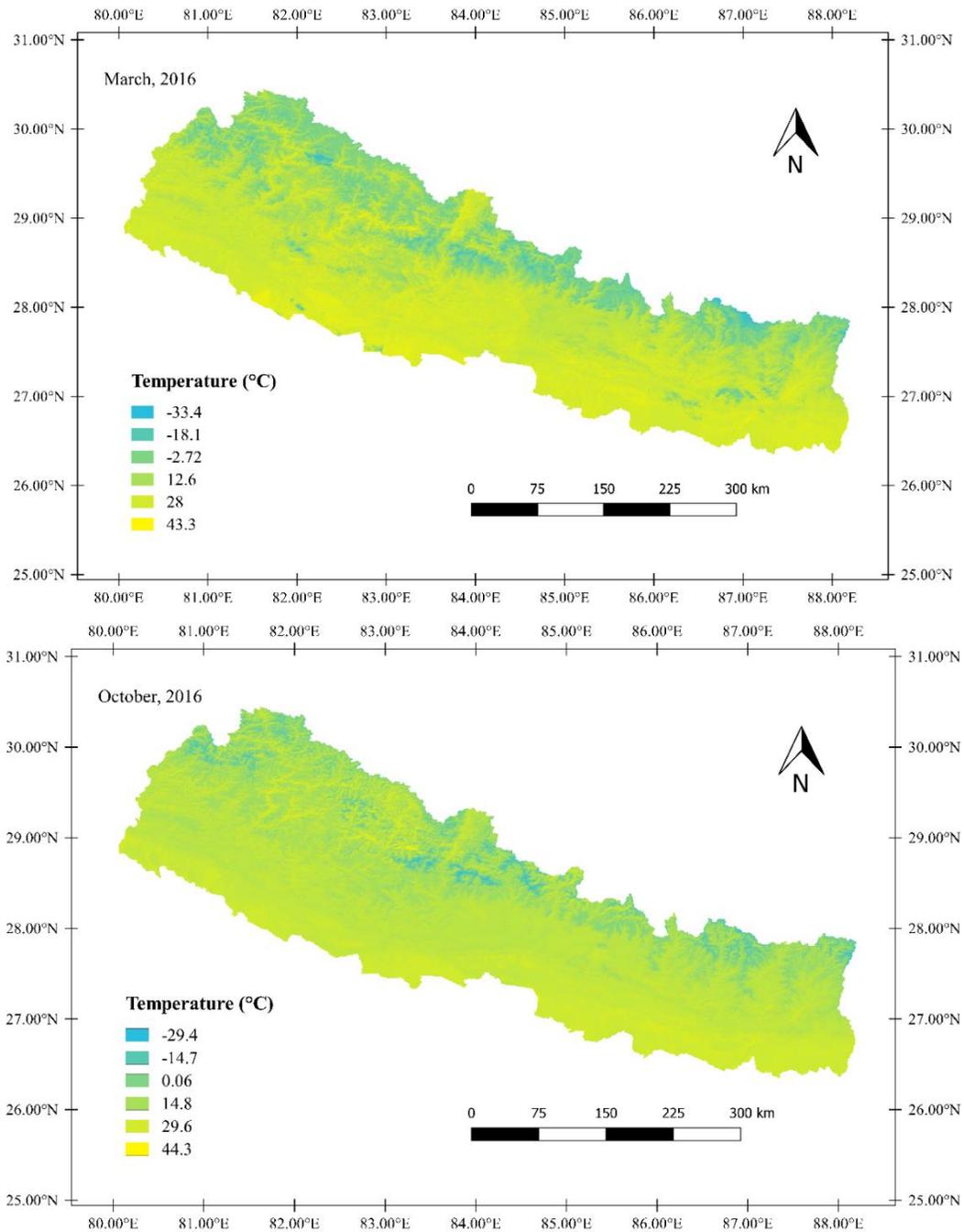

**Figure 9** METRIC derived T$_s$ for March 2016 (upper) and, October 2016 (lower) at the time of satellite overpass.

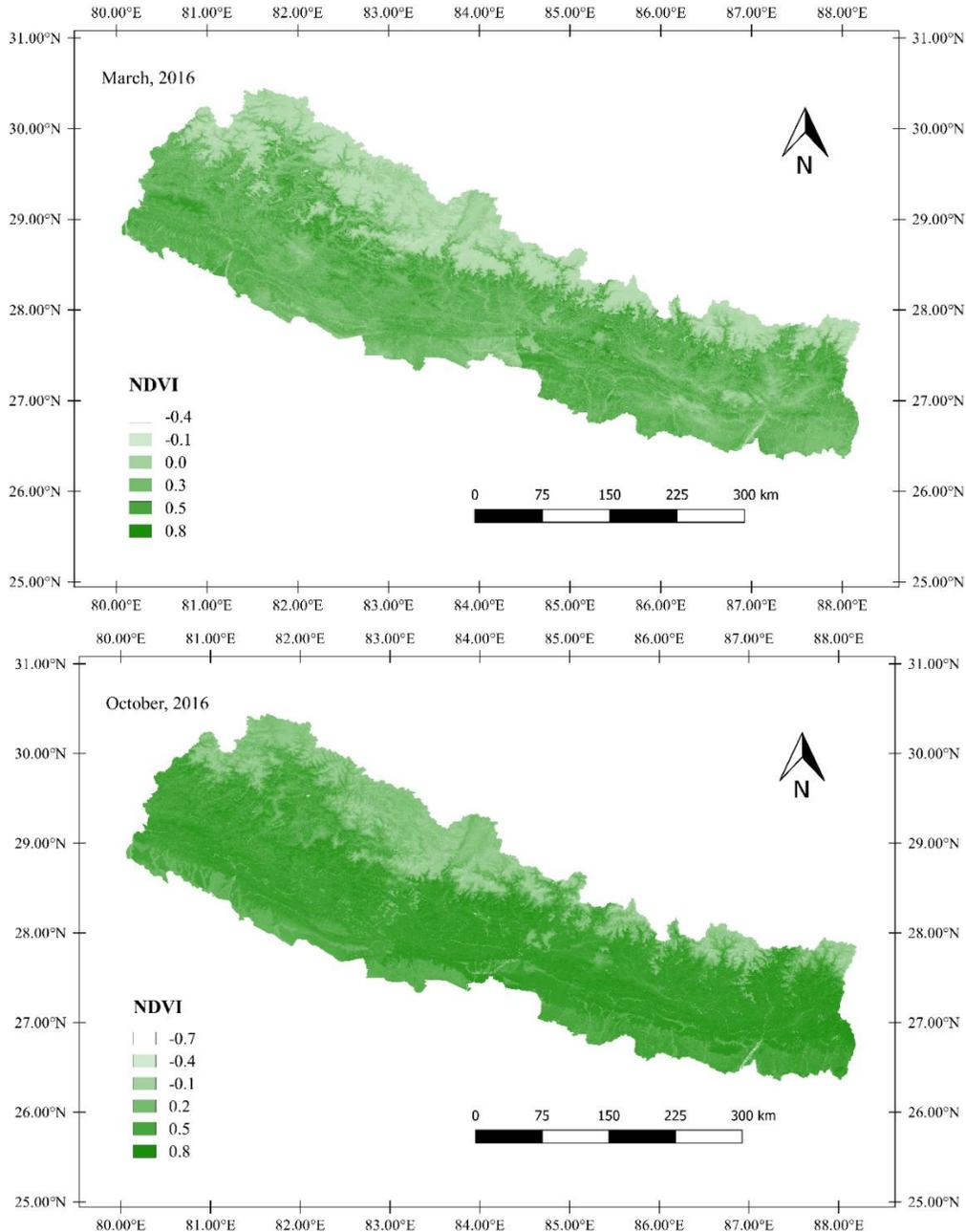

**Figure 10** METRIC derived NDVI for March 2016 (upper) and, October 2016 (lower) at the time of satellite overpass.

After the computation of all the intermediate variables, ET at the time of satellite overpass was computed as a final output from the METRIC model. We computed ET for the entire region of Nepal in the month of March and October and the results obtained are shown in **Figure 11**. For March, minimum ET obtained is 0.002 mm/hr and the maximum was 1.15 mm/hr. While for the month of October, the minimum was 0.002 mm/hr and the maximum was 1.2 mm/hr. The average ET for March and October was obtained as 0.373 mm/hr and 0.393 mm/hr respectively. ET at the time of satellite overpass for October is high as compared to March.

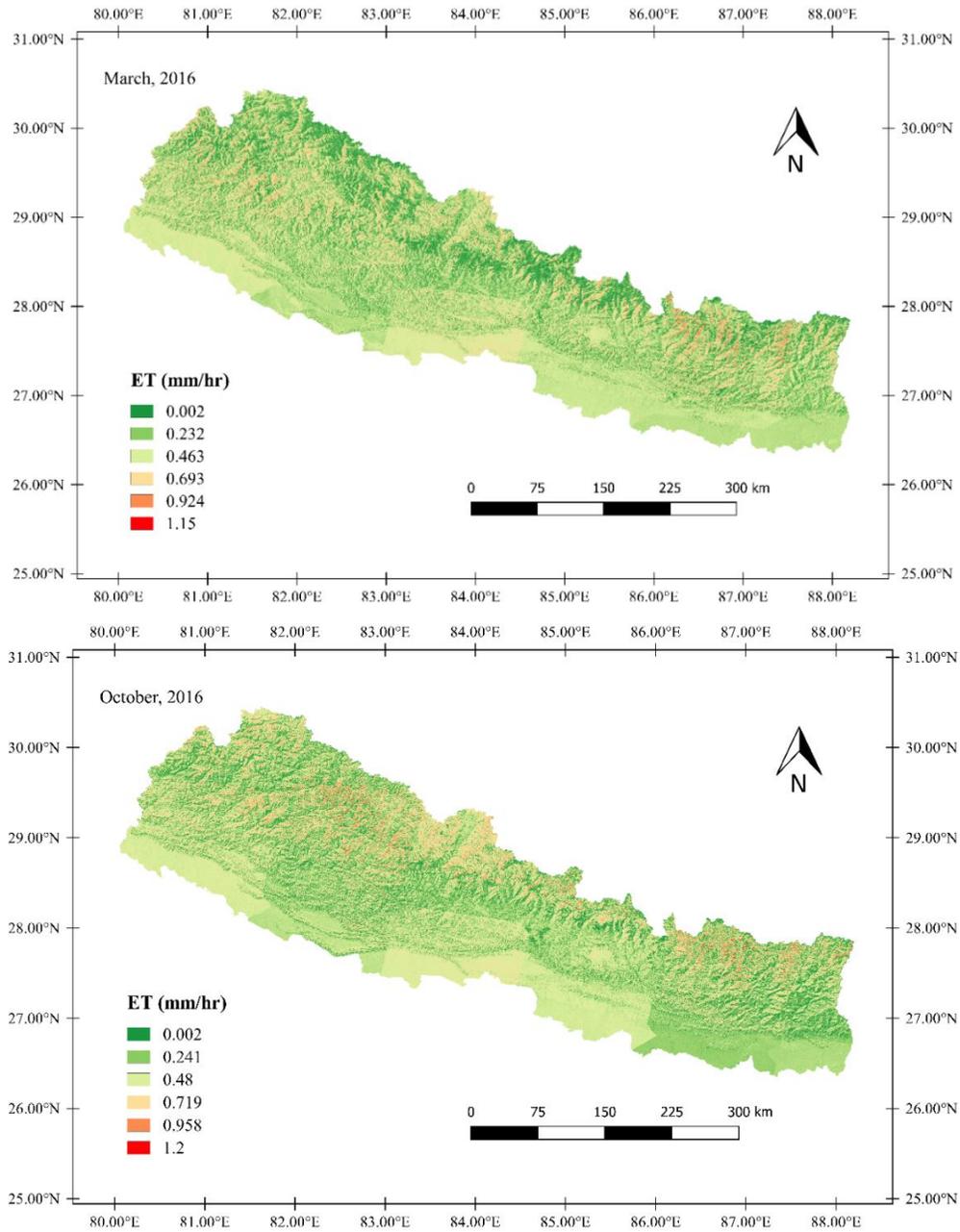

**Figure 11** METRIC derived ET at the time of satellite overpass for March 2016 (upper) and, October 2016 (lower)

## 3.4 Discussion

Computation of ET has applications in several areas such as agriculture monitoring, climate forecasting, etc. Previous studies concerning ET computation for a topographically diversified region like Nepal have used only low-resolution remote sensing solutions. In this work, we investigated ET estimation using high-resolution imagery from Landsat 8. We used the METRIC model for ET computation which requires the calculation of different energy components. These were computed as intermediate variables in the METRIC model. A large variation from North to South is to be expected in our study area for different energy components, given the variations in the topology that occurs from North to South. Guided by the availability of cloud-free Landsat 8 images for 2016, we computed heat fluxes, vegetation indices, surface albedo, $T_s$, and ET for March and October for the entire study area (Nepal).

We validated our obtained results of $T_s$ and ET estimation with the EC station data from Simara. We see that our results from the METRIC model are close with the weather station results indicating that the METRIC model is suitable for ET estimation in topographically complex terrain as our study area. We obtained an RMSE of 0.06 mm/hr and 1.24 mm/day for ET estimation. In previous works (as discussed in Chapter 1), we see that the METRIC model has been applied in different areas from plain flat lands to high mountainous regions. We compare our obtained RMSE for ET estimation with other results discussed in Chapter 1, though these results are from different regions. In a study conducted over an irrigated field of Saudi Arabia by (Madugundu et al., 2017) using the METRIC model and eddy covariance (EC) flux tower data, the authors got an RMSE of 0.13 mm/hr and 4.15 mm/day for ET estimation. ET estimates using the METRIC model in South Dakota by (Reyes-González et al., 2017) resulted in an RMSE of 0.65 mm/day. He et al. (2017) estimates ET in an Almond orchard field of California using the METRIC model. An RMSE of 0.11mm/hr in all seasons and 0.09 mm/hr in growing seasons were obtained. Comparing our RMSE for ET estimation with the RMSE reported in these other studies, we can assess that our results from the model are generally good.

We computed the variations in $T_s$ and NDVI for the entire region of Nepal. The results of $T_s$ for March and October are shown in **Figure 9**. As it can be seen in the figure, variation can be seen in $T_s$ with changes in elevation. As the elevation increases, $T_s$ decreases for both the months. The average $T_s$ for the complete study area of Nepal for March and October was obtained as 18.695 ˚C, and 19.112 ˚C. Temperature is observed to be low at the beginning of pre-monsoon season, and of moderate value in the post-monsoon season; March is a pre-monsoon season in Nepal and October is considered as the post-monsoon season (Nayava, 1980). These observations are as expected and establish a basic validation of our computation. It is to be noted that in our model we use $T_s$ also for the condition selection of hot and cold pixel required for calculation of H. Variation is also seen in NDVI as computed from the METRIC model for the entire study area as shown in **Figure 10**. In the lower elevation of the study area, NDVI is seen to be high and in the higher elevation, NDVI is almost seen as zero. Lower NDVI at the higher elevation of Nepal is likely because higher elevation areas comprise snow and glaciers.

After the computation of all the intermediate features, we computed ET at the time of satellite overpass for March and October corresponding to our entire study area, with an elevation range from 57 m to 8848 m, and is shown in **Figure 11**. Average ET for March was obtained as 0.373 mm/hr with an elevation-wise variation of ET (high in lower elevation and low in higher elevation) for our study area. In October, ET seems to be high with an average of 0.393 mm/hr as compared to March. Higher ET for the month of October is due to the higher temperatures and higher NDVI for this compared, in comparison to March, as was observed in the results of the values for $T_s$ and NDVI.

While we discussed the computation and validation of ET for the region of Nepal, with the model validated we also analyze patterns seen with ET over the areas of Nepal. In particular we study what sort of variations are seen in ET with changing elevation. This is described in the next Chapter.

# Chapter 4 Elevation wise variation of ET

## 4.1 Background

Different parameters affect ET such as $T_s$, vegetation and the topography of the area. Within a small region from north to south of Nepal, there lies an agricultural area, forest, shrublands, water bodies, grasslands, barren areas, snow, and glaciers. Due to the mixed landcover in this area, variations in ET with changes in elevation can be expected. In Chapter 2, we presented the METRIC model as a remote sensing methodology to enable large scale ET estimation. In our work, we considered Landsat 8 based images for processing by METRIC in order to estimate ET for the region of Nepal. In Chapter 3, we presented the results of ET estimation for Nepal and validation. By comparing with the ground station measurements, we found that METRIC based ET estimation using Landsat 8 matches closely with weather station-based measurements. Thus, remote sensing can be used for large scale ET estimation even for a complex topographical region like Nepal.

Since the METRIC model is capable to provide good ET estimation, it is further interesting to observe the elevation-wise variation of ET in Nepal based on ET estimated from the METRIC model and identify any trends across different months. Our selected region of Nepal provides a good study area for the elevation-wise variation of ET analysis because the elevation range of Nepal is large. The elevation changes from 57 m to 8848 m even though the span of the country for this elevation range is only 193 km on average. While observing and analyzing the elevation wise variations of ET, it might be interesting to also investigate how $T_s$ and NDVI change with elevation. Land surface temperature represented by $T_s$ is an important parameter related to ET through evaporation relation mainly. Similarly, NDVI which relates to the vegetation in the area has an impact on ET mostly through the transpiration relation. Besides, $T_s$ and NDVI there are numerous other factors that impact ET. Therefore, through direct relationship can be expected of ET and $T_s$ or NDVI, $T_s$ and NDVI alone cannot probably account for observed variations in ET as per the elevation changes.

## 4.2 Method

To see the elevation wise variation for a different month, and as per the availability of the cloud-free Landsat 8 data for 2016, we computed ET of different months (March, April, May, June, October, November) for a portion of the study area. For this, we take 12 Landsat 8 image tiles corresponding to the path and row as (141, 41) and (141,40). This region covers the entire North-South region of Nepal to study elevation wise variations. The area we select to see the elevation wise variation on ET is shown in **Figure 12**. Also, the selected region incorporates both the weather station that we have used in our work here.

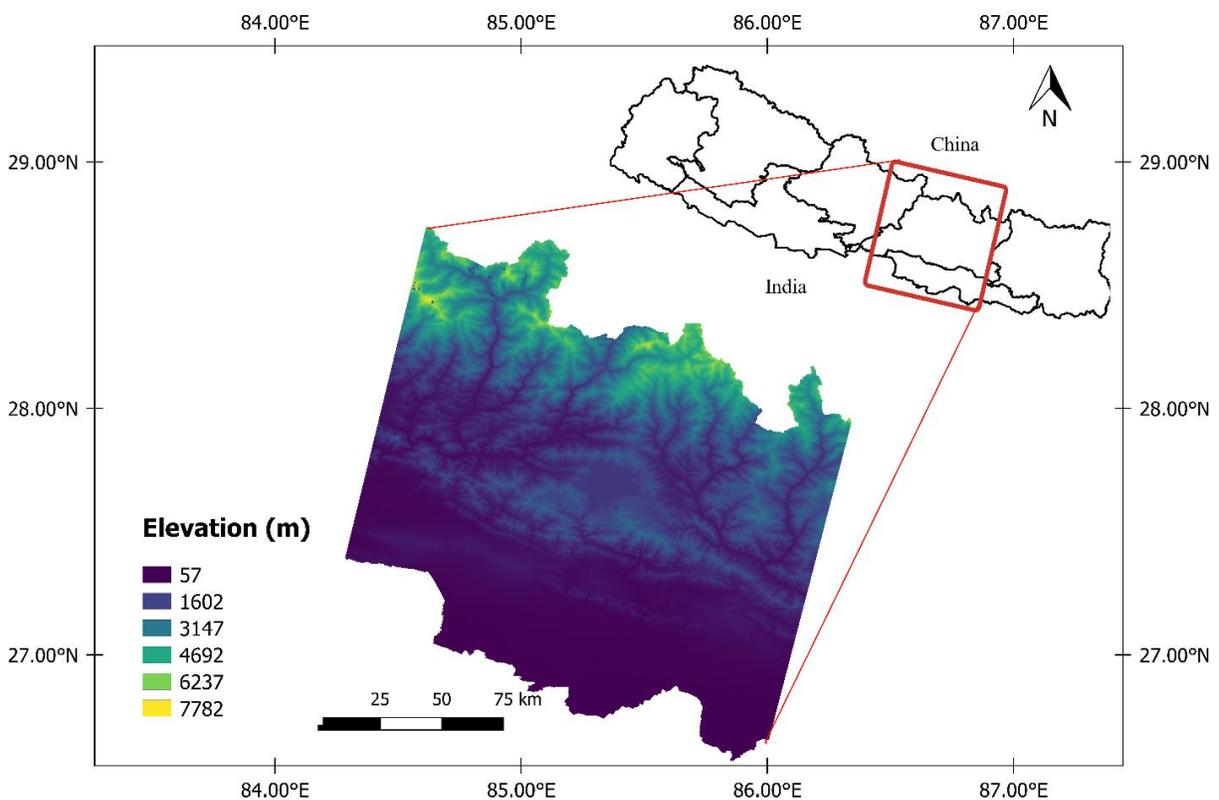

**Figure 12** Study area selected to see the elevation wise variations. The color map shows different elevation distribution.

## 4.3 Results

### 4.3.1 Elevation-wise variation of Intermediate results derived from METRIC model

Using our METRIC model implementation, $T_s$, vegetation indices, surface albedo, and heat fluxes are all computed as intermediate products. To see the elevation wise variation on ET, we first computed $T_s$, and NDVI for six different months as $T_s$, and NDVI has a direct impact on ET. The graphical representation of elevation wise variation on NDVI is shown in **Figure 13**. In all the months we saw that NDVI is high in the lower elevation and low in the higher elevation. $T_s$ was also plotted against elevation to see the variation on $T_s$ as per elevation as shown in **Figure 14**. We see that $T_s$ is high in the lower elevation and low in higher elevation for all the months. The average value for $T_s$, NDVI and all the heat fluxes obtained for the selected region are summarized in **Table 6**.

**Table 6** METRIC derived intermediate products for month March, April, May, June, October and November 2016

| Months | avg $T_s$ (˚C) | avg NDVI | avg $R_n$ (Wm⁻²) | avg G (Wm⁻²) | avg H (Wm⁻²) | avg LE (Wm⁻²) |
|---|---|---|---|---|---|---|
| March | 19.1 | 0.36 | 610.83 | 92.13 | 292.56 | 226.11 |
| April | 23.62 | 0.21 | 608.86 | 94.23 | 345.75 | 168.66 |
| May | 24.48 | 0.41 | 655.75 | 100.74 | 326.28 | 228.73 |
| June | 22.12 | 0.34 | 353.73 | 59.64 | 166.57 | 127.05 |
| October | 20.72 | 0.51 | 680.62 | 91.21 | 336.92 | 249.91 |
| November | 20.48 | 0.52 | 420.59 | 60.15 | 200.76 | 157.39 |

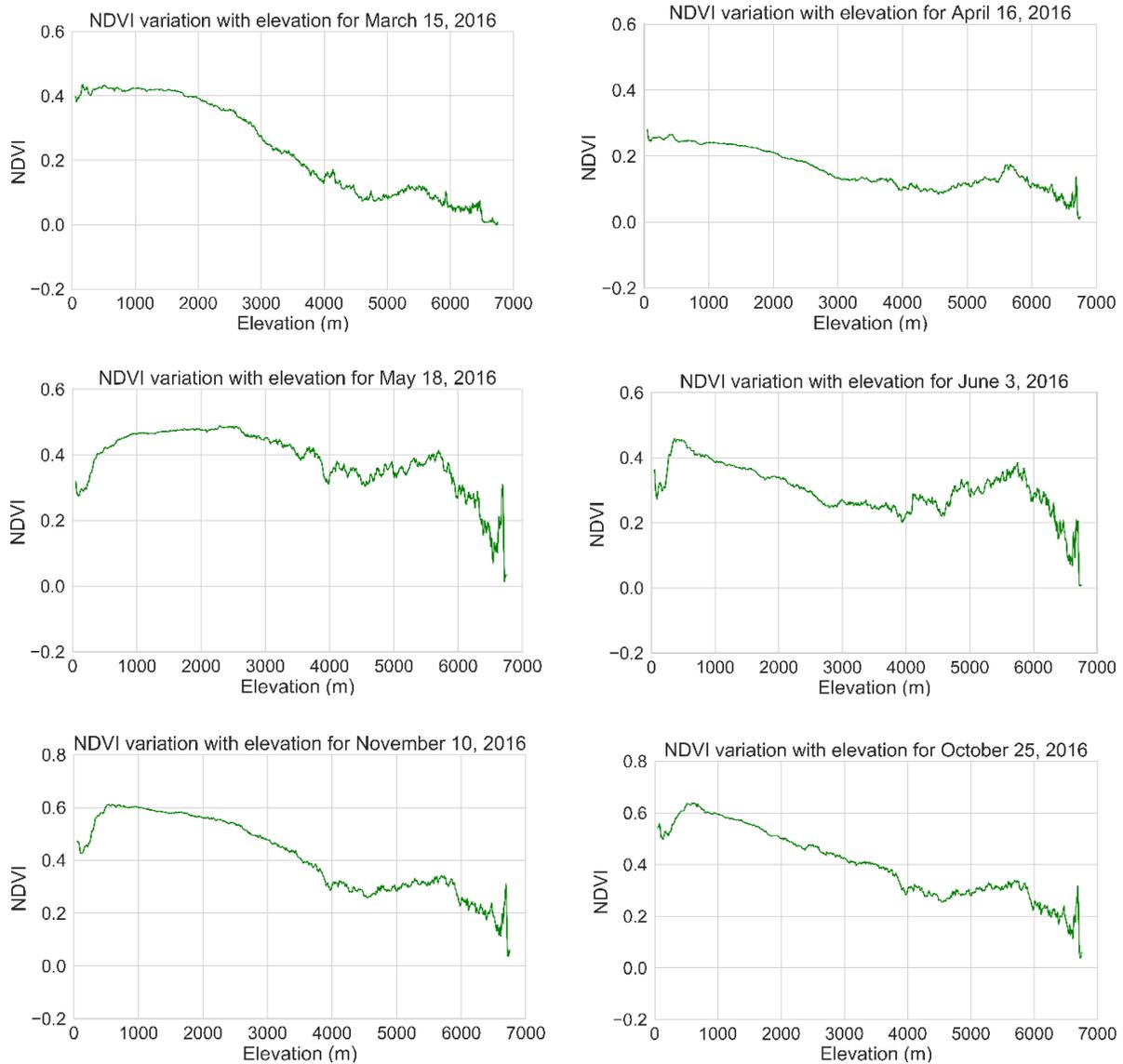

**Figure 13** NDVI at the time of satellite overpass for March 15, 2016 (upper left), April 16, 2016 (upper right) May 18, 2016 (middle left), June 3, 2016 (middle right), October 25, 2016 (lower left) and, November 10, 2016 (lower right).

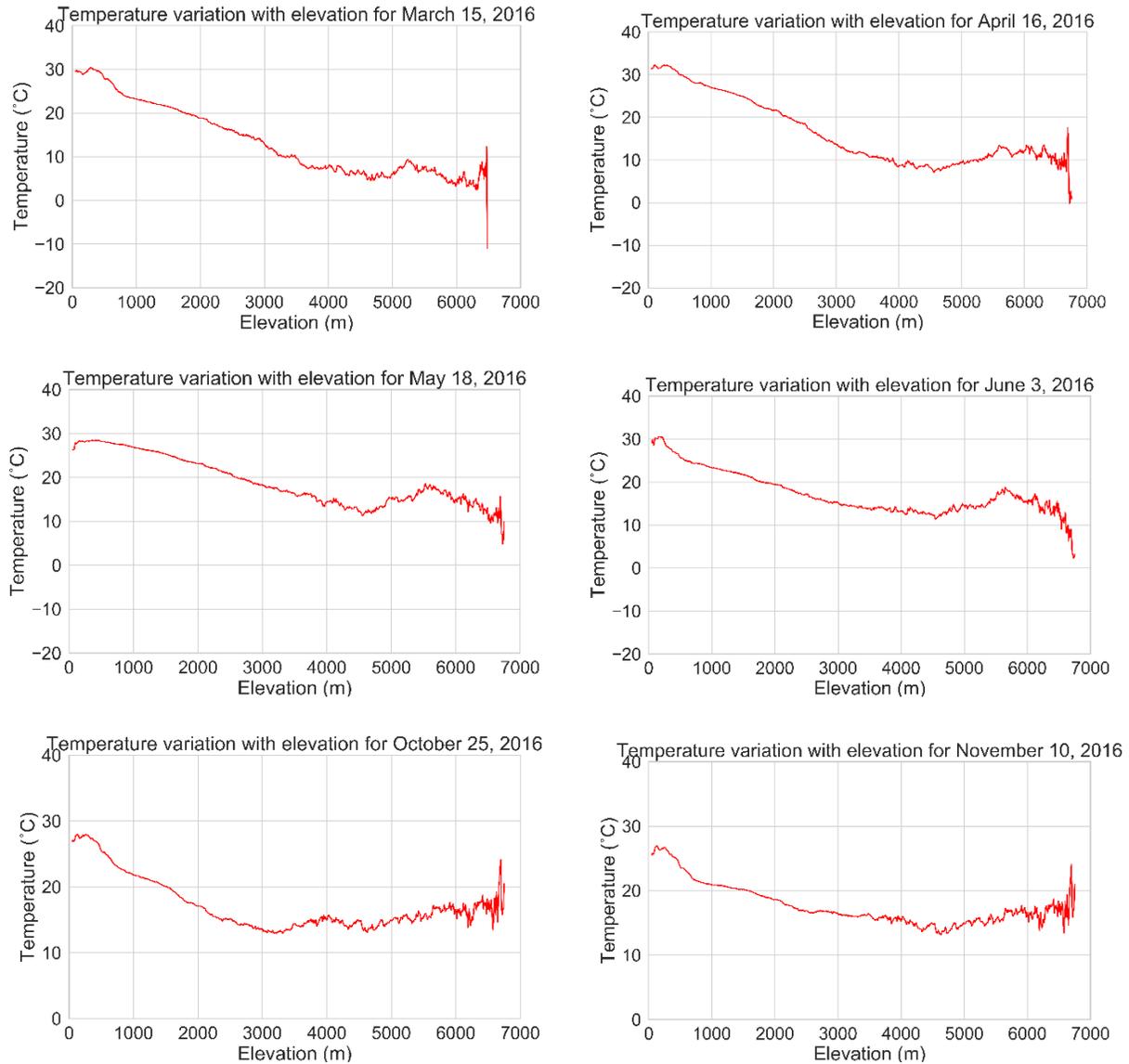

**Figure 14** $T_s$ at the time of satellite overpass for March 15, 2016 (upper left), April 16, 2016 (upper right) May 18, 2016 (middle left), June 3, 2016 (middle right), October 25, 2016 (lower left) and, November 10, 2016 (lower right).

### 4.3.2 Elevation-wise variation of Hourly and daily ET derived from the METRIC model

Based on the computation of all the heat fluxes, vegetation indices, and $T_s$, $ET_{inst}$ was computed. After the computation of the ET at the time of satellite overpass, daily ET was computed from $ET_{inst}$ and reference ET ($ET_r$) obtained from in-situ data.

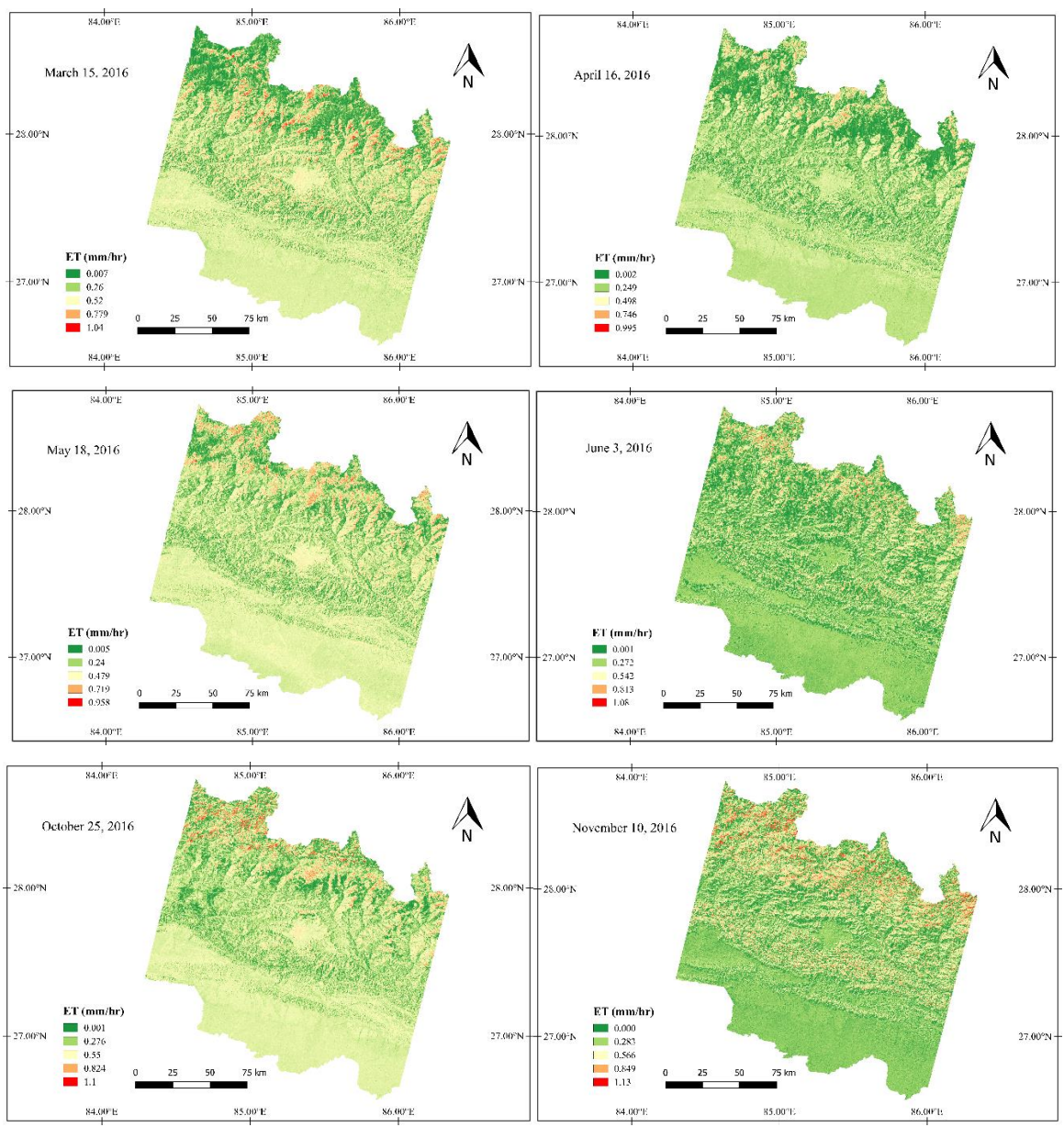

**Figure 15** MERTIC derived ET at the time of satellite overpass for March 15, 2016 (upper left), April 16, 2016 (upper right) May 18, 2016 (middle left), June 3, 2016 (middle right), October 25, 2016 (lower left) and, November 10, 2016 (lower right).

Avg $ET_{inst}$ obtained from the METRIC model for the month March, April, May, June, October, and November was 0.36 mm/hr, 0.28 mm/hr, 0.35 mm/hr, 0.26 mm/hr, 0.39 mm/hr, and 0.31 mm/hr respectively while daily ET was obtained as 4.33 mm/day, 7.009 mm/day, 4.29 mm/day, 3.24 mm/day, 3.97 mm/day, and 1.5 mm/day respectively. ET at the time of satellite overpass for different months is shown in **Figure 15**.

To have a clear insight on ET, the obtained ET for six different months was graphically plotted against elevation as shown in **Figure 16**. ET showed variations with changes in elevation. In the majority of the months, ET decreases as the elevation increases. In June and November, ET increased to a certain elevation and again decreases as the elevation increased.

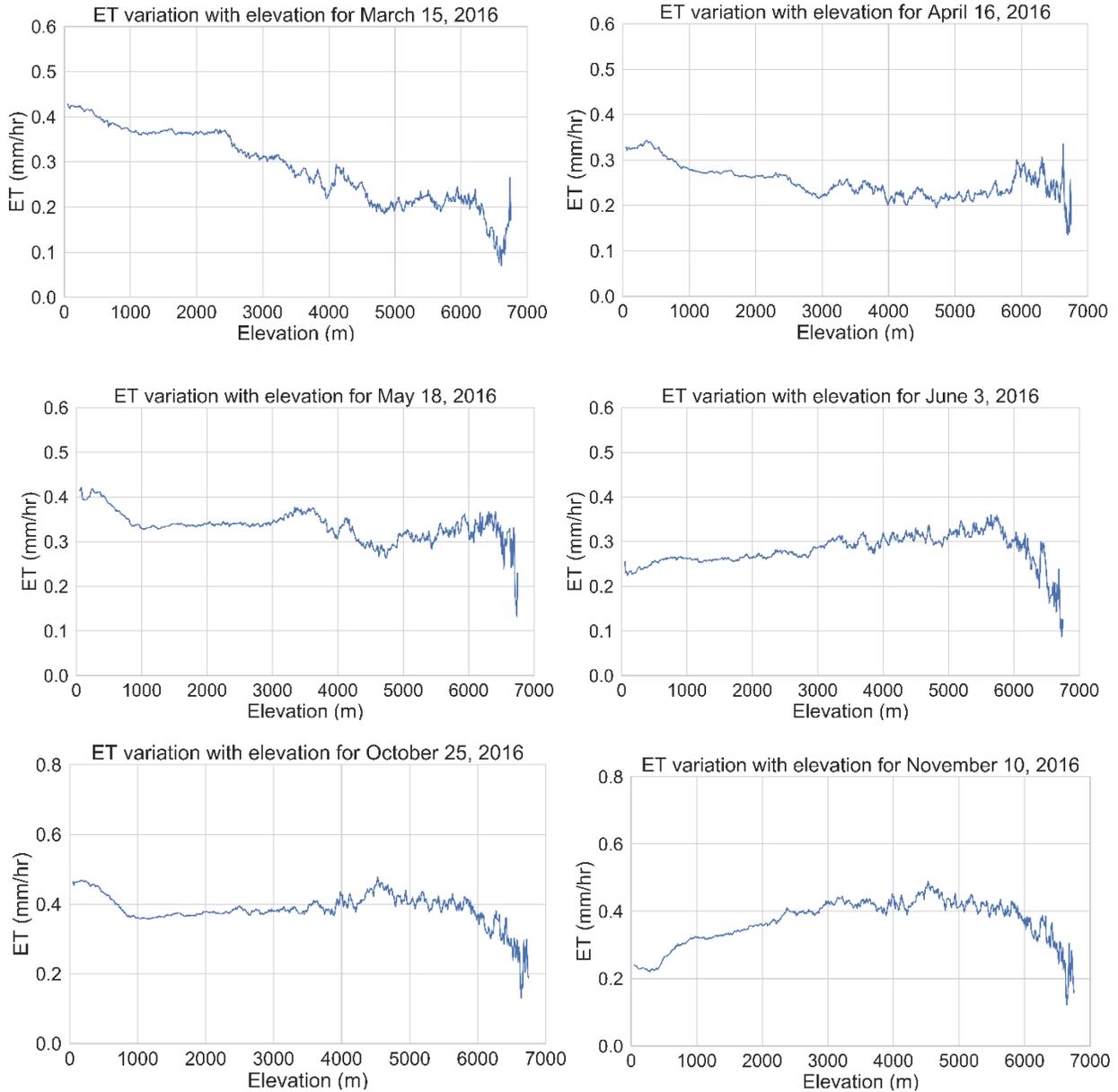

**Figure 16** Variation of ET with elevation for March 15, 2016 (upper left), April 16, 2016 (upper right) May 18, 2016 (middle left), June 3, 2016 (middle right), October 25, 2016 (lower left) and, November 10, 2016 (lower right).

### 4.3.3 Variation of ET, T$_s$, and NDVI with elevation

We plotted the estimated ET, T$_s$, and NDVI compared to the corresponding elevation for our selected sub-region. The plots show the elevation-wise variations in ET, T$_s$, and NDVI for six different months as given in **Figure 17**. In March, April, and May, maximum ET was obtained as 0.42 mm/hr, 0.33 mm/hr, and 0.41 mm/hr while minimum ET was obtained as 0.07 mm/hr, 0.14 mm/hr, and 0.13 mm/hr respectively. In March, April, and May maximum ET was seen in the lower elevation between 57m to 300 m and minimum ET above 6000 m. For June, October, and November, in the lower elevation below 300m, ET was obtained as 0.26 mm/hr, 0.47 mm/hr, and 0.24 mm/hr respectively. Maximum ET obtained for June, October, and November was 0.36 mm/hr, 0.48 mm/hr, and 0.49 and minimum ET is 0.09 mm/hr, 0.13 mm/hr, and 0.12 mm/hr respectively. T$_s$ and NDVI were seen to be high in the lower elevation and low in the higher elevation. Maximum T$_s$ obtained for March, April, May, June, October, and November was 30˚C, 32 ˚C, 29 ˚C, 30.5 ˚C, 27 ˚C, and 26.5 ˚C and minimum T$_s$ was -13 ˚C, -6 ˚C, 3 ˚C, 4 ˚C, -1 ˚C, and -1 ˚C respectively. For March, April, May, June, October, and November, maximum NDVI is obtained as 0.4, 0.28, 0.43, 0.48, 0.64, and 0.61 and minimum NDVI as -0.02, 0.0, -0.01, -0.02, -0.08, and -0.07 respectively.

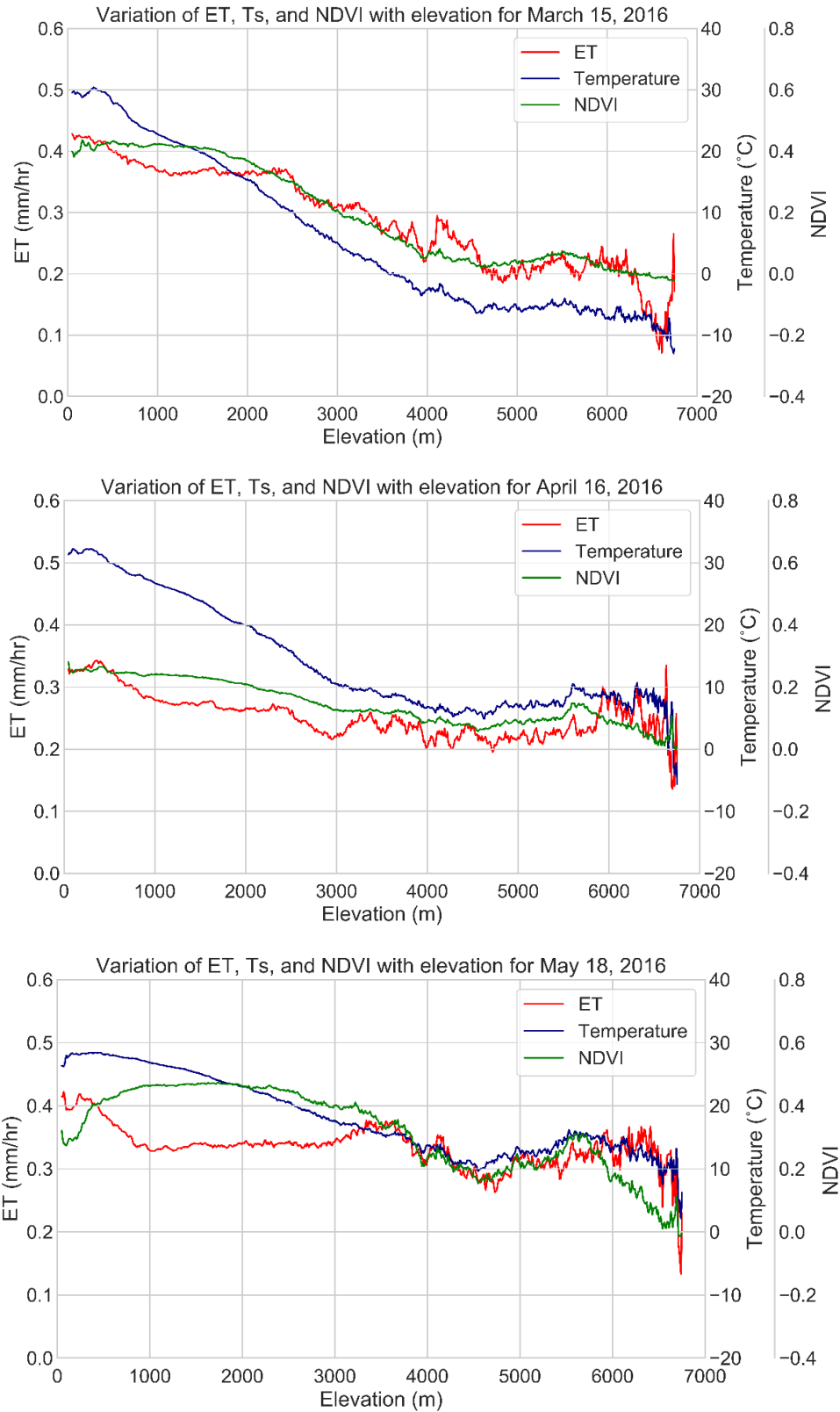

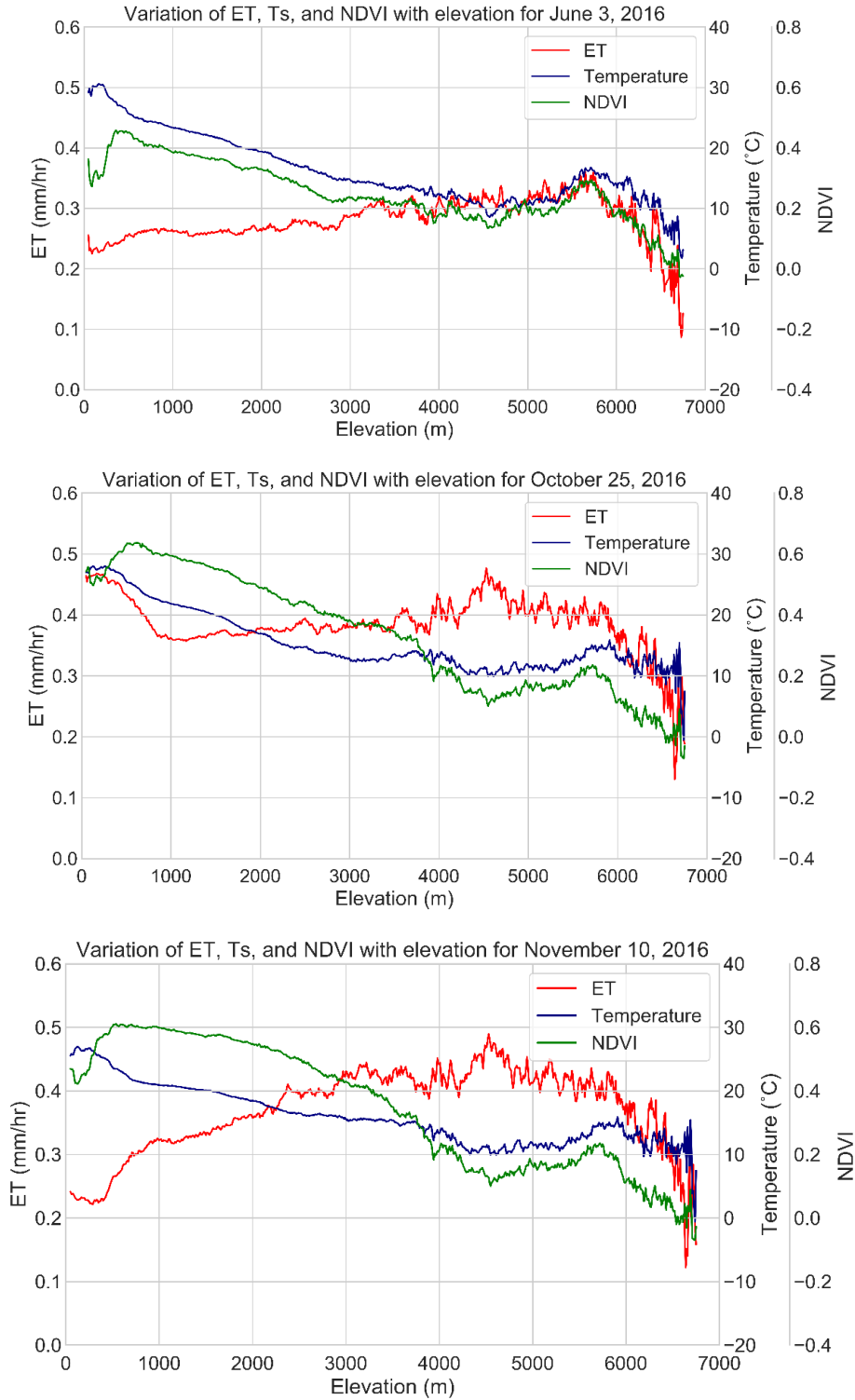

**Figure 17** Variation of ET, T$_s$, and NDVI with elevation for March 15, 2016 (upper left), April 16, 2016 (upper right) May 18, 2016 (middle left), June 3, 2016 (middle right), October 25, 2016 (lower left) and, November 10, 2016 (lower right).

## 4.4 Discussion

We also aimed to analyze the elevation wise variation of ET and to compute daily ET. For this, we took a sub-region within our study area where more cloud-free Landsat 8 images were available. We took 12 Landsat 8 images for path row (141, 41) and (141,40), covering the entire North to South in the central region of Nepal, to analyze elevation wise variation on ET for different months (March, April, May, June, October, and November). The selected region also covered our both of the ground-based weather station used for calibration and validation.

To see the elevation wise variation on ET for different months, we computed ET on a portion of our study area for six different months. The selection of a smaller study area covering the north-south region provides representative elevation for the whole Nepal region while avoiding the problem of having many tiles covered with the cloud when the whole region would have to be considered for analysis. We selected the central region of our study area that covers the south to north area and includes both weather stations used in our study. We computed all the intermediate parameters ($T_s$, NDVI, SAVI, LAI, surface albedo, $R_n$, G, H, and LE) for six different months and the average value for these parameters is shown in **Table 6**. A similar pattern in the $T_s$ and NDVI comparison across months, i.e. higher $T_s$ and NDVI in October as compared to March, was seen in this selected sub-region also as we had earlier obtained based on the analysis of the whole region. This shows the validation of our assumptions that the selected North-south region is representative for the entire Nepal region. After getting all the parameters, ET at the time of satellite overpass and daily ET is computed. The ET at the time of satellite overpass for March, April, May, June, October, and November months which has been used to study elevation-wise variations of ET is shown in **Figure 15**.

We plotted the obtained ET, $T_s$, and NDVI for different months with elevation to observe the elevation wise variation of ET which is shown in **Figure 17**. In March, April, and May, it is observed that the ET decreases as the elevation increases. Lower elevation region of our study area mostly comprises of agricultural areas and higher elevation consists of barren areas and snow/glaciers (**Figure 6**). In these three different months, ET in the lower elevation is seen to be high and it gradually falls off to the higher elevation. We see a similar trend for $T_s$, and NDVI also. Thus, higher ET at a lower elevation is generally due to higher temperatures and vegetation present there. In Nepal, as one goes towards Northwards, both general reduction in temperature and vegetation happens. This leads to the general trend of reducing ET with elevation. However, ET is a complex parameter that depends not only on land surface temperature and vegetation but also on numerous other parameters like wind speed, land terrain properties, etc. This leads to some aberrations seen in the general trend in elevation-wise variation of ET for some elevation ranges e.g. in the month of November. Considering the aberrations in ET and elevation relation as seen in the month of November, one likely explanation for ET not decreasing with elevation is because the $T_s$ and NDVI do not decrease as sharply in this month. Therefore, other parameters that affect ET could have likely more effect to dwarf the effect of $T_s$ and NDVI on ET. Referring to our results reported for all months, in the lower elevations ET is high as lower elevation is mainly covered by vegetations. By referring to the land-use map also, we could see that lower elevation within a range from 57 m to 1000 m, mostly consists of agricultural areas where irrigation is also high which implies a higher ET.

# Chapter 5 Conclusion and Recommendation

## Conclusion

Topographical diversity of regions like Nepal requires ET estimation with fine spatial resolution, especially given the lack of ground station infrastructures to cover large areas. In this work, we have implemented and evaluated a remote sensing-based estimation of ET utilizing the high-resolution Landsat 8 data using the METRIC model. ET and $T_s$ estimated from our model were found to be in close agreements with the field measurements obtained from an independent EC flux tower data. We computed ET and other climate parameters across two months corresponding to pre-monsoon seasons and post-monsoon seasons, for whole Nepal. The obtained trend across the month matches what would be expected, e.g. higher temperature in the post-monsoon season for Nepal compared to the pre-monsoon season given our selected months. This further confirms that remote sensing methodology can be useful for trend analysis across a large region. An advantage of remote sensing-based estimation is that multiple climate parameters can be estimated from the same source of information. This, for example, allowed us simultaneously to estimate $T_s$ and NDVI besides ET. We observed the direct effect that $T_s$ and vegetation have on ET. METRIC model, in particular, has modeling steps accounting for topological variations like the presence of mountains and varying slopes. This makes the model highly applicable for regions like Nepal, as is evident from our results too.

Remote sensing-based ET estimation allows to study relations like elevation wise variation of ET. Such a study is suited for our selected area, Nepal, which has large variations in elevation. A study on the relation of ET with elevation is not only interesting scientifically to better understand the dynamics of ET but also has highly practical applications such as for agriculture planning. On investigating the elevation-wise variation in ET, we found an inverse relation between elevation and ET for our considered region in Nepal probably due to high vegetations of lower altitudes and glacier/snow-capped mountains in the higher altitudes.

Our work is the first to study and validate the use of high-resolution satellite imagery for ET estimation in the context of the topographical region like Nepal. Large scale estimation of ET with the fine spatial resolution has numerous applications for a country like Nepal, e.g. in the context of proper agricultural planning and forecasting.

## Recommendation

The model performance obtained with our METRIC implementation compared favorably with other results reported in the literature for ET estimation. This showed that the METRIC model is already capable to provide a good estimate of climate parameters such as ET to be relevant for use in assessment and planning for a region like Nepal. However, several improvements could still be made in the model and analysis of the estimated ET can be extended.

Currently, we use an automatic method for the estimation of cold and hot pixels based on land use maps and surface temperature. It must be further investigated how this pixel selection criteria fares with other automatic pixel selection methods and manual pixel selection approaches. Similarly, we did not address the pixels likely corrupted by the presence of cloud except when searching for the anchor pixels related to hot and cold pixel selection. The presence of the cloud will likely impact the final ET estimation and needs to be addressed separately. One approach would be to interpolate results corresponding to any pixels likely affected by the presence of clouds based on neighboring cloud-free pixels. In our results, the presence of cloud was deemed to be not much of an issue as we selected only the images with less cloud cover. However, if the ET estimation is to be scaled to be useful for all the months and time period, a mechanism to deal with the presence of cloud cover must be implemented.

Further, we validated our ET estimation based on only one ground-based station data currently. In the future work, subject to the availability of such data, measurements from different ground station data at different elevation and topographic locations should be used for model validation.

Being the first study of its kind for the region of Nepal, in this work, we only considered preliminary analysis of ET patterns over Nepal, e.g. studying how ET varies with elevation. Future works should consider further analysis such as understanding ET variations across land-use types. Similarly, it would also be interesting to observe the trend of ET variations over the years. Though many people in Nepal are still relying on agriculture, scientific farming is not adopted widely yet leading to poor outputs. Based on our work of ET estimation, a possible extension towards use in an application would be to collaborate with the agricultural planning department in Nepal and consider adapting farming practices based on estimated ET and other climatic conditions.

# References

Allen, R., Hartogensis, O., de Bruin, H., 2000. Long-wave radiation over alfafa during the RAPID field campaign in southern Idaho. Research Report, Kimberly, Univ. of Idaho, Id.

Allen, R., Walter, I., Elliot, R., Howell, T., Itenfisu, D., Jensen, M., 2005. The ASCE standardized reference evapotranspiration equation. ASCE-EWRI task committee final report.

Allen, R.G., Pereira, L.S., Raes, D., Smith, M., 1998. Crop evapotranspiration-Guidelines for computing crop water requirements-FAO Irrigation and drainage paper 56. Fao, Rome 300, D05109.

Allen, R.G., Tasumi, M., Trezza, R., 2007. Satellite-based energy balance for mapping evapotranspiration with internalized calibration (METRIC)—Model. Journal of irrigation and drainage engineering 133, 380-394.

Allen, R.G., Trezza, R., Kilic, A., Tasumi, M., Li, H., 2013. Sensitivity of Landsat-scale energy balance to aerodynamic variability in mountains and complex terrain. JAWRA Journal of the American Water Resources Association 49, 592-604.

Amatya, P.M., Ma, Y., Han, C., Wang, B., Devkota, L.P., 2015. Recent trends (2003–2013) of land surface heat fluxes on the southern side of the central Himalayas, Nepal. Journal of Geophysical Research: Atmospheres 120, 11,957-911,970.

Amatya, P.M., Ma, Y., Han, C., Wang, B., Devkota, L.P., 2016. Mapping regional distribution of land surface heat fluxes on the southern side of the central Himalayas using TESEBS. Theoretical and applied climatology 124, 835-846.

Baral, T., 2012. Evapotransiration from Natural and Planted Forest in the Middle Mountain of Nepal. University of Twente Faculty of Geo-Information and Earth Observation (ITC).

Bastiaanssen, W., Noordman, E., Pelgrum, H., Davids, G., Thoreson, B., Allen, R., 2005. SEBAL model with remotely sensed data to improve water-resources management under actual field conditions. Journal of irrigation and drainage engineering 131, 85-93.

Bastiaanssen, W.G., 1998. Remote sensing in water resources management: The state of the art. International Water Management Institute.

Bastiaanssen, W.G., 2000. SEBAL-based sensible and latent heat fluxes in the irrigated Gediz Basin, Turkey. Journal of hydrology 229, 87-100.

Bastiaanssen, W.G., Menenti, M., Feddes, R., Holtslag, A., 1998a. A remote sensing surface energy balance algorithm for land (SEBAL). 1. Formulation. Journal of hydrology 212, 198-212.

Bastiaanssen, W.G., Pelgrum, H., Wang, J., Ma, Y., Moreno, J., Roerink, G., Van der Wal, T., 1998b. A remote sensing surface energy balance algorithm for land (SEBAL).: Part 2: Validation. Journal of hydrology 212, 213-229.

Bastiaanssen, W.G.M., 1995. Regionalization of surface flux densities and moisture indicators in composite terrain: A remote sensing approach under clear skies in Mediterranean climates. SC-DLO.

Brunsell, N.A., Gillies, R.R., 2002. Incorporating surface emissivity into a thermal atmospheric correction. Photogrammetric engineering and remote sensing 68, 1263-1270.

Brutsaert, W., 1982. Evaporation into the atmosphere: Theory, history, and applications. D. Reidel Publ., Boston, MA. Evaporation into the atmosphere: Theory, history, and applications. D. Reidel Publ., Boston, MA., -.

Carlson, T.N., Ripley, D.A., 1997. On the relation between NDVI, fractional vegetation cover, and leaf area index. Remote sensing of Environment 62, 241-252.

Chaudhary, P., Aryal, K.P., 2009. Global warming in Nepal: challenges and policy imperatives. Journal of Forest and livelihood 8, 5-14.

Chavez, J., Gowda, P., Evett, S., Colaizzi, P., Howell, T., Marek, T., 2007. An application of METRIC for ET mapping in the Texas high plains. Trans ASABE (submitted).

Choudhury, B.J., DiGirolamo, N.E., 1998. A biophysical process-based estimate of global land surface evaporation using satellite and ancillary data I. Model description and comparison with observations. Journal of Hydrology 205, 164-185.

Cooper, P., 1969. The absorption of radiation in solar stills. Solar energy 12, 333-346.

Dahal, P., Shrestha, N.S., Shrestha, M.L., Krakauer, N.Y., Panthi, J., Pradhanang, S.M., Jha, A., Lakhankar, T., 2016. Drought risk assessment in central Nepal: temporal and spatial analysis. Natural Hazards 80, 1913-1932.

Dirmeyer, P.A., Gao, X., Zhao, M., Guo, Z., Oki, T., Hanasaki, N., 2006. The Second Global Soil Wetness Project (GSWP-2): Multi-model analysis and implications for our perception of the land surface. Bull. Am. Meteorol. Soc 87, 1381-1397.

Duffie, J.A., Beckman, W.A., 2013. Solar engineering of thermal processes. John Wiley & Sons.

Dungey, J.W., 1961. Interplanetary magnetic field and the auroral zones. Physical Review Letters 6, 47.

Engman, E.T., Gurney, R.J., 1991. Remote sensing in hydrology. Chapman and Hall London.

Fisher, J.B., Melton, F., Middleton, E., Hain, C., Anderson, M., Allen, R., McCabe, M.F., Hook, S., Baldocchi, D., Townsend, P.A., 2017. The future of evapotranspiration: Global requirements for ecosystem functioning, carbon and climate feedbacks, agricultural management, and water resources. Water Resources Research 53, 2618-2626.

Folhes, M., Rennó, C., Soares, J., 2009. Remote sensing for irrigation water management in the semi-arid Northeast of Brazil. Agricultural Water Management 96, 1398-1408.

Friedlingstein, P., Meinshausen, M., Arora, V.K., Jones, C.D., Anav, A., Liddicoat, S.K., Knutti, R., 2014. Uncertainties in CMIP5 climate projections due to carbon cycle feedbacks. Journal of Climate 27, 511-526.

Fritts, H., 2012. Tree rings and climate. Elsevier.

Garrison, J.D., Adler, G.P., 1990. Estimation of precipitable water over the United States for application to the division of solar radiation into its direct and diffuse components. Solar Energy 44, 225-241.

George, B.A., Reddy, B., Raghuwanshi, N., Wallender, W., 2002. Decision support system for estimating reference evapotranspiration. Journal of Irrigation and Drainage Engineering 128, 1-10.

Ghimire, C.P., Lubczynski, M.W., Bruijnzeel, L.A., Chavarro-Rincón, D., 2014. Transpiration and canopy conductance of two contrasting forest types in the Lesser Himalaya of Central Nepal. Agricultural and forest meteorology 197, 76-90.

Goulden, M.L., Bales, R.C., 2014. Mountain runoff vulnerability to increased evapotranspiration with vegetation expansion. Proceedings of the National Academy of Sciences 111, 14071-14075.

Han, C., Ma, Y., Chen, X., Su, Z., 2017. Trends of land surface heat fluxes on the Tibetan Plateau from 2001 to 2012. International journal of climatology 37, 4757-4767.

He, R., Jin, Y., Kandelous, M.M., Zaccaria, D., Sanden, B.L., Snyder, R.L., Jiang, J., Hopmans, J.W., 2017. Evapotranspiration estimate over an almond orchard using landsat satellite observations. Remote Sensing 9, 436.

Healey, N.C., Irmak, A., Arkebauer, T.J., Billesbach, D.P., Lenters, J.D., Hubbard, K.G., Allen, R.G., Kjaersgaard, J., 2011. Remote sensing and in situ-based estimates of evapotranspiration for subirrigated meadow, dry valley, and upland dune ecosystems in the semi-arid sand hills of Nebraska, USA. Irrigation and drainage systems 25, 151-178.

Hipps, L.E., 1989. The infrared emissivities of soil and Artemisia tridentata and subsequent temperature corrections in a shrub-steppe ecosystem. Remote sensing of environment 27, 337-342.

Huete, A.R., 1988. A soil-adjusted vegetation index (SAVI). Remote sensing of environment 25, 295-309.

Idso, S.B., Jackson, R.D., Reginato, R.J., 1975. Estimating evaporation: a technique adaptable to remote sensing. Science 189, 991-992.

IPCC, 2007. summary for policymakers, climate change 2007: the physical science basis. Contribution of Working Group I to the Fourth Assessment Report of the Intergovernmental Panel on Climate Change." United Kingdom and New York, USA (2007). Climate Change.

Irmak, A., Ratcliffe, I., Ranade, P., Hubbard, K.G., Singh, R.K., Kamble, B., Kjaersgaard, J., 2011. Estimation of land surface evapotranspiration with a satellite remote sensing procedure. Great plains research, 73-88.

Joshi, B.B., Ma, Y., Ma, W., Sigdel, M., Wang, B., Subba, S., 2020. Seasonal and diurnal variations of carbon dioxide and energy fluxes over three land cover types of Nepal. Theoretical and Applied Climatology 139, 415-430.

Jung, M., Reichstein, M., Ciais, P., Seneviratne, S.I., Sheffield, J., Goulden, M.L., Bonan, G., Cescatti, A., Chen, J., De Jeu, R., 2010. Recent decline in the global land evapotranspiration trend due to limited moisture supply. Nature 467, 951.

Kalma, J.D., McVicar, T.R., McCabe, M.F., 2008. Estimating Land Surface Evaporation: A Review of Methods Using Remotely Sensed Surface Temperature Data. Surveys in Geophysics 29, 421-469.

Koster, R.D., Dirmeyer, P.A., Guo, Z., Bonan, G., Chan, E., Cox, P., Gordon, C., Kanae, S., Kowalczyk, E., Lawrence, D., 2004. Regions of strong coupling between soil moisture and precipitation. Science 305, 1138-1140.

Lambert, L., Chitrakar, B., 1989. Variation of potential evapotranspiration with elevation in Nepal. Mountain Research and Development, 145-152.

Lawrence, D.M., Oleson, K.W., Flanner, M.G., Thornton, P.E., Swenson, S.C., Lawrence, P.J., Zeng, X., Yang, Z.L., Levis, S., Sakaguchi, K., 2011. Parameterization improvements and functional and structural advances in version 4 of the Community Land Model. Journal of Advances in Modeling Earth Systems 3.

Lian, J., Huang, M., 2016. Comparison of three remote sensing based models to estimate evapotranspiration in an oasis-desert region. Agricultural Water Management 165, 153-162.

Liang, S., 2001. Narrowband to broadband conversions of land surface albedo I: Algorithms. Remote sensing of environment 76, 213-238.

Liebert, R., Huntington, J., Morton, C., Sueki, S., Acharya, K., 2016. Reduced evapotranspiration from leaf beetle induced tamarisk defoliation in the Lower Virgin River using satellite-based energy balance. Ecohydrology 9, 179-193.

Ma, W., Hafeez, M., Ishikawa, H., Ma, Y., 2013. Evaluation of SEBS for estimation of actual evapotranspiration using ASTER satellite data for irrigation areas of Australia. Theoretical and applied climatology 112, 609-616.

Ma, W., Hafeez, M., Rabbani, U., Ishikawa, H., Ma, Y., 2012. Retrieved actual ET using SEBS model from Landsat-5 TM data for irrigation area of Australia. Atmospheric environment 59, 408-414.

Madugundu, R., Al-Gaadi, K.A., Tola, E., Hassaballa, A.A., Patil, V.C., 2017. Performance of the METRIC model in estimating evapotranspiration fluxes over an irrigated field in Saudi Arabia using Landsat-8 images. Hydrology and Earth System Sciences 21, 6135.

Mahto, S.S., Pandey, A.C., 2018. Satellite Based Temporal Analysis of Local Weather Elements along N–S Transect across Jharkhand, Bihar and Eastern Nepal, Multidisciplinary Digital Publishing Institute Proceedings, 343.

Mancosu, N., Snyder, R., Kyriakakis, G., Spano, D., 2015. Water Scarcity and Future Challenges for Food Production. Water, 7, 975-992. DOI 10, w7030975.

Menenti, M., 1993. Parameteraization of land surface evaporation by means of location dependent potential evaporation and surface temperature range. Bolle, Feddes and Kalma, Exchange processes at the land surface for a range of space and time scales.

Meng, X., Evans, J., McCabe, M., 2014. The impact of observed vegetation changes on land–atmosphere feedbacks during drought. Journal of Hydrometeorology 15, 759-776.

Mielnick, P., Dugas, W., Mitchell, K., Havstad, K., 2005. Long-term measurements of CO2 flux and evapotranspiration in a Chihuahuan desert grassland. Journal of Arid environments 60, 423-436.

Miralles, D., De Jeu, R., Gash, J., Holmes, T., Dolman, A., 2011. Magnitude and variability of land evaporation and its componentsat the global scale. Hydrology and Earth System Sciences.

Miralles, D.G., Van Den Berg, M.J., Gash, J.H., Parinussa, R.M., De Jeu, R.A., Beck, H.E., Holmes, T.R., Jiménez, C., Verhoest, N.E., Dorigo, W.A., 2014. El Niño–La Niña cycle and recent trends in continental evaporation. Nature Climate Change 4, 122.

Naegeli, K., Damm, A., Huss, M., Wulf, H., Schaepman, M., Hoelzle, M., 2017. Cross-Comparison of albedo products for glacier surfaces derived from airborne and satellite (Sentinel-2 and Landsat 8) optical data. Remote Sensing 9, 110.

Nayava, J.L., 1980. Rainfall in Nepal. Himalayan Review 12, 1-18.

Norman, J.M., Becker, F., 1995. Terminology in thermal infrared remote sensing of natural surfaces. Remote Sensing Reviews 12, 159-173.

Numata, I., Khand, K., Kjaersgaard, J., Cochrane, M.A., Silva, S.S., 2017. Evaluation of Landsat-based METRIC modeling to provide high-spatial resolution evapotranspiration estimates for Amazonian forests. Remote Sensing 9, 46.

Reyes-González, A., Kjaersgaard, J., Trooien, T., Hay, C., Ahiablame, L., 2017. Comparative Analysis of METRIC Model and Atmometer Methods for Estimating Actual Evapotranspiration. International Journal of Agronomy 2017.

Roerink, G., Su, Z., Menenti, M., 2000. S-SEBI: A simple remote sensing algorithm to estimate the surface energy balance. Physics and Chemistry of the Earth, Part B: Hydrology, Oceans and Atmosphere 25, 147-157.

Seneviratne, S.I., Lüthi, D., Litschi, M., Schär, C., 2006. Land–atmosphere coupling and climate change in Europe. Nature 443, 205.

Sobrino, J., Raissouni, N., 2000. Toward remote sensing methods for land cover dynamic monitoring: Application to Morocco. International journal of remote sensing 21, 353-366.

Sobrino, J.A., Jiménez-Muñoz, J.C., Paolini, L., 2004. Land surface temperature retrieval from LANDSAT TM 5. Remote Sensing of environment 90, 434-440.

Song, L., Zhuang, Q., Yin, Y., Zhu, X., Wu, S., 2017. Spatio-temporal dynamics of evapotranspiration on the Tibetan Plateau from 2000 to 2010. Environmental Research Letters 12, 014011.

Stathopoulou, M., Cartalis, C., 2007. Daytime urban heat islands from Landsat ETM+ and Corine land cover data: An application to major cities in Greece. Solar Energy 81, 358-368.

Su, Z., 2002. The Surface Energy Balance System (SEBS) for estimation of turbulent heat fluxes. Hydrology and earth system sciences 6, 85-100.

Suzuki, R., Masuda, K., Dye, D.G., 2007. Interannual covariability between actual evapotranspiration and PAL and GIMMS NDVIs of northern Asia. Remote Sensing of Environment 106, 387-398.

Tasumi, M., 2003. Progress in operational estimation of regional evapotranspiration using satellite imagery. University of Idaho.

Tsouni, A., Kontoes, C., Koutsoyiannis, D., Elias, P., Mamassis, N., 2008. Estimation of actual evapotranspiration by remote sensing: Application in Thessaly Plain, Greece. Sensors 8, 3586-3600.

Uddin, K., Shrestha, H.L., Murthy, M., Bajracharya, B., Shrestha, B., Gilani, H., Pradhan, S., Dangol, B., 2015. Development of 2010 national land cover database for the Nepal. Journal of environmental management 148, 82-90.

Valipour, M., 2017. Analysis of potential evapotranspiration using limited weather data. Applied Water Science 7, 187-197.

Vautard, R., Yiou, P., D'andrea, F., De Noblet, N., Viovy, N., Cassou, C., Polcher, J., Ciais, P., Kageyama, M., Fan, Y., 2007. Summertime European heat and drought waves induced by wintertime Mediterranean rainfall deficit. Geophysical Research Letters 34.

Wagle, P., Gowda, P.H., 2019. Editorial for the Special Issue "Remote Sensing of Evapotranspiration (ET)". Multidisciplinary Digital Publishing Institute.

Wang, K., Dickinson, R.E., Wild, M., Liang, S., 2010. Evidence for decadal variation in global terrestrial evapotranspiration between 1982 and 2002: 1. Model development. Journal of Geophysical Research: Atmospheres 115.

Wang, S.-Y., Yoon, J.-H., Gillies, R.R., Cho, C., 2013. What caused the winter drought in western Nepal during recent years? Journal of Climate 26, 8241-8256.

Wild, M., Folini, D., Schär, C., Loeb, N., Dutton, E.G., König-Langlo, G., 2013. The global energy balance from a surface perspective. Climate dynamics 40, 3107-3134.

Zhang, H., Anderson, R.G., Wang, D., 2015a. Satellite-based crop coefficient and regional water use estimates for Hawaiian sugarcane. Field Crops Research 180, 143-154.

Zhang, K., Kimball, J.S., Nemani, R.R., Running, S.W., Hong, Y., Gourley, J.J., Yu, Z., 2015b. Vegetation greening and climate change promote multidecadal rises of global land evapotranspiration. Scientific reports 5, 15956.

Zhang, K., Kimball, J.S., Running, S.W., 2016a. A review of remote sensing based actual evapotranspiration estimation. Wiley Interdisciplinary Reviews: Water 3, 834-853.

Zhang, Y., Peña-Arancibia, J.L., McVicar, T.R., Chiew, F.H., Vaze, J., Liu, C., Lu, X., Zheng, H., Wang, Y., Liu, Y.Y., 2016b. Multi-decadal trends in global terrestrial evapotranspiration and its components. Scientific reports 6, 19124.